\documentclass[conference]{IEEEtran}
\IEEEoverridecommandlockouts
\usepackage{cite}
\usepackage{amsmath,amssymb,amsfonts}
\usepackage{algorithmic}
\usepackage{graphicx}
\usepackage{textcomp}
\usepackage{array}
\usepackage[table]{xcolor}
\usepackage[hyphens]{url}
\usepackage{booktabs}
\newcommand{\system}{\textsc{TaskArtisan}}
\newcommand{\numpar}{12}
\usepackage{csquotes}
\usepackage{tabularx}
\usepackage{caption}
\usepackage{enumitem}
\usepackage{multirow} 

\def\BibTeX{{\rm B\kern-.05em{\sc i\kern-.025em b}\kern-.08em
T\kern-.1667em\lower.7ex\hbox{E}\kern-.125emX}}
\begin{document}

\title{\system{}: Designing Composable Generative Widgets for LLM-Assisted Analysis}


\author{
\IEEEauthorblockN{Meng Chen, Amy Pavel}
\IEEEauthorblockA{
Department of EECS, University of California, Berkeley, CA, USA \\
\{meng.chen, amypavel\}@berkeley.edu
}
}

\maketitle

\begin{abstract}
People increasingly use chatbots such as ChatGPT for everyday analysis tasks. While chatbots unify many analysis functions (e.g., scripts, visualizations, summaries), long conversations become hard to navigate, making it difficult to revisit prior steps or reuse successful workflows. LLMs now generate high-fidelity GUI code that enables people to create customized analysis tools beyond text. Yet, what new opportunities generative UIs bring to analysis work remain unclear. We interviewed six professionals about analysis with chatbots, analyzed publicly shared LLM-generated GUI tools, and conducted a comparison study (N=12) between a chatbot and \system{}, a technology probe that enables people to create and assemble generative analysis UI widgets for sequential and fan-out composition. We find that GUI improved clarity and visual presentation but also introduced rigidity and additional prompting challenges. We summarize the trade-offs into a provisional design framework (\textit{malleability, specification, interoperability}) to inform future generative UI in LLM-assisted analysis workflows.
\end{abstract}


\section{Introduction}
People analyze data (\textit{e.g.}, text, spreadsheets)~\cite{gu_how_2024, zhang_how_2020} to surface results (\textit{e.g.}, summaries, visualizations) and ultimately make decisions (\textit{e.g.}, select a college, plan inventory)~\cite{yun_generative_2025, fok_marco_2024, noauthor_passages_nodate}.
Traditionally, people use multiple applications to complete such analyses~\cite{yun_generative_2025, cao_generative_2025}. For example, a college counselor might use a browser to look up school information, then use Excel to plot expected cost against financial aid, and finally use Google Docs to present the pros, cons, and images of each school in a table.
New Large Language Model (LLM) chatbots like ChatGPT Analyst~\cite{noauthor_data_nodate} and Claude Artifact~\cite{noauthor_claude_nodate} unify diverse analysis and visualization capabilities into one application. Using these chatbots, people can upload their \textbf{data} then write a prompt to describe what \textbf{analysis} they want to perform. Thus, they can complete an entire analysis workflow in a single application (\textit{e.g.}, create a plot then a comparison table). However, while reuse is often central to data analysis workflows (\textit{e.g.}, with Computational Notebooks, shared scripts), iterative analyses with LLM chatbots remain static and embedded in long chat conversations~\cite{suh_sensecape_2023,zhang_visar_2023,freund_flowco_2025} such that it remains difficult to adjust or reuse prior analysis work~\cite{payandeh_noteex_2025, rule_exploration_2018, noauthor_managing_nodate}. 

\begin{figure}[htb]
  \centering
  \includegraphics[width=3.33in]{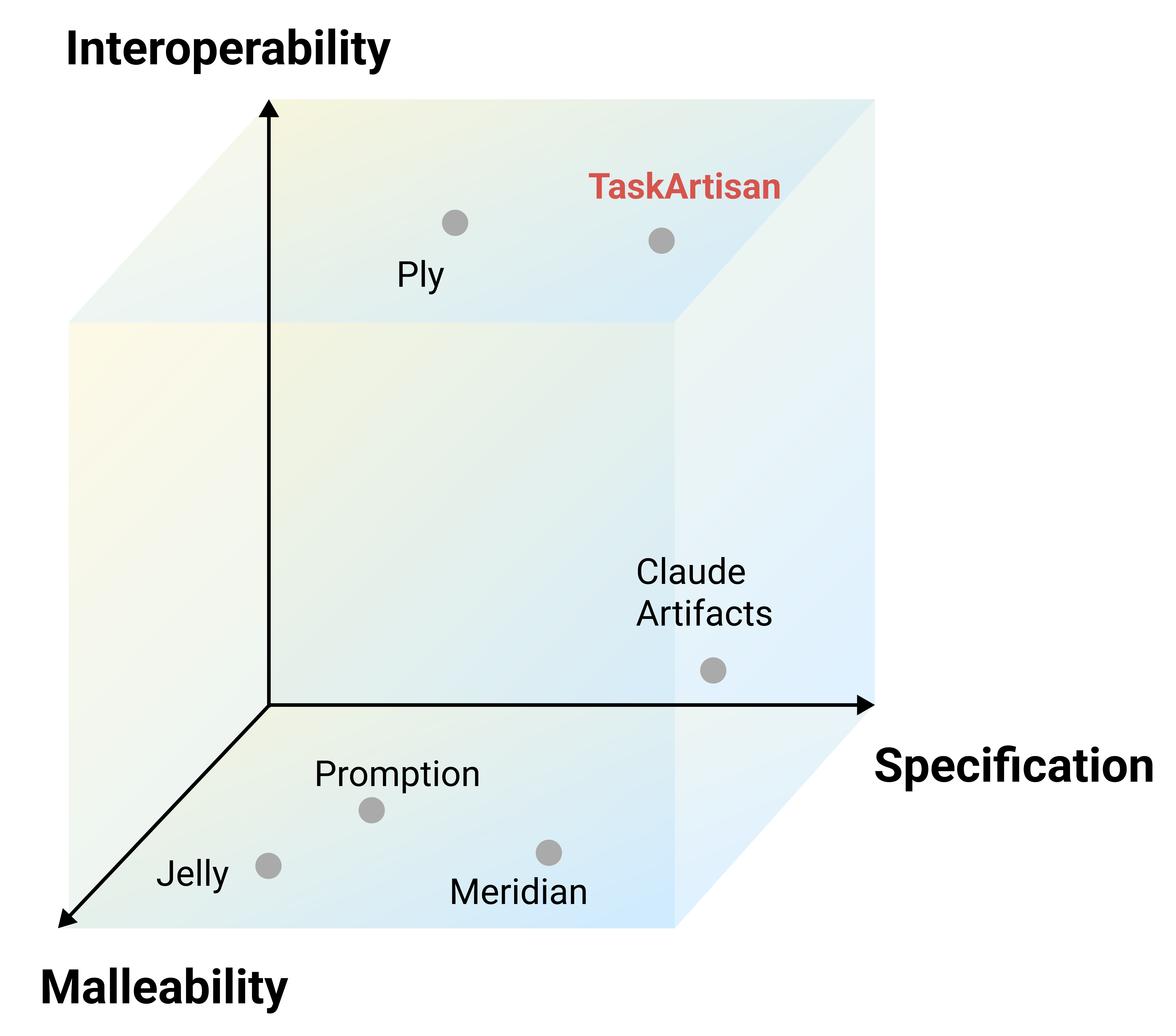}
  \caption{Three-axis design framework for generative UI tools: malleability after generation, degree of user specification, and interoperability between UIs. Existing systems occupy different regions: Jelly~\cite{cao_generative_2025} and Meridian~\cite{min_meridian_2025} emphasize malleable modification; Promption~\cite{drosos_dynamic_2025} provides options in UIs for ambiguous open-ended prompts; Ply~\cite{aveni_generative_2025} prioritizes interoperability. \system{} targets high interoperability and malleability for reusable analysis workflows, but fuzzy analytical tasks that evolve during data inspection point to a design opportunity for generative UIs that adapt to emerging needs with less upfront specification.}
  \label{fig:framework}
\end{figure}

Now, LLMs let people generate interactive Graphical User Interfaces (GUIs) by writing text prompts with applications such as Claude Artifacts~\cite{noauthor_claude_nodate} or Replit~\cite{noauthor_replit_nodate}. 
As such prompt-to-GUI applications become widely accessible, there is a potential for people to create functional and reusable LLM-generated GUIs for their common analysis tasks (\textit{e.g.}, adjust student preferences on the UI and output a standardized comparison table), thus offering the power and flexibility of LLM-based analysis in a reusable form. Prior HCI work explored embodying prompts as GUI to support reuse of commands and facilitate non-linear iterative workflows in writing~\cite{noauthor_cells_nodate} and image editing~\cite{ai_instrument}.
We explore the opportunity to use generative UIs for analysis through three interrelated research questions: \textit{(i) What are the existing opportunities and practices for using generative UIs for analysis?}, \textit{(ii)  What are trade-offs between using chat and generative UIs for analysis?}, and \textit{(iii) What are design opportunities to improve the usefulness of generative UIs for analysis?}

To understand how professionals might use LLM-generated GUIs in day-to-day work, we first interviewed 6 professionals about how they currently use LLMs and what automated support they want in their analysis workflows. As all 6 professionals currently used LLM chatbots rather than LLM-generated GUIs, we then asked the same participants to use the publicly available Claude Artifacts tool to create an LLM-generated GUI and reflect on the benefits and challenges of current LLM-generated GUIs compared to LLM chatbots. To understand the broader current use of LLM-generated GUIs in analysis work, we analyzed LLM-generated GUIs in the Data Analysis category on Claude Artifacts website and annotated each GUI’s data, output, interactivity, and analysis goal. 
We found that while people recognized the importance of and already developed workarounds (\textit{e.g.,} reuse one conversations window to maintain prior context) to reuse common analyses, each individual LLM-generated GUI \textbf{tightly couples the data, analysis, and UI together} such that reuse on different data remains non-trivial~\cite{rule_exploration_2018, gadhave_persist_2024, muller_how_2019}. Classic literature has emphasized the importance of customization and reuse in software design (\textit{e.g.}, Model-View-Controller) ~\cite{mackay_patterns_1990,deacon_model-view-controller_nodate}, which is still missing in current LLM-generated GUIs.

To explore how to address current limitations of LLM-generated GUIs for analysis work, we developed \system{}, a technology probe~\cite{hutchinson_technology_2003} that augments current LLM-generated GUIs in three ways. First, to comply with current main usage of LLM in analysis work, we defined a generative \textbf{widget} as an LLM-generated GUI that has a LLM component to analyze data input and present the result on a GUI component. Second, to make widgets reusable, we decouple the \textbf{data input}, \textbf{LLM component}, and \textbf{GUI component} so LLM analysis can be reused across data. Last, as real analysis often involves sequences of steps and various visualizations~\cite{rule_exploration_2018}, we introduce \textbf{sequential} and \textbf{fan-out} composition so people can use \system{} to compose multiple generative widgets into repeatable workflows and/or dashboards.

To assess the trade-off between our system and state-of-the-art LLM chatbots with generative visualization, we conducted a controlled study with \numpar{} professionals comparing \system{} to Claude Artifacts. 
While participants spent a similar amount of time completing a set of analysis tasks with both interfaces, participants faced a clear trade-off: participants spent more time authoring their \textit{initial analysis} workflow when using \system{} than when using Claude Artifacts, but participants completed \textit{repeated analysis} tasks faster when using \system{} than when using Claude Artifacts. 
Participants attributed reuse benefits to \system{}'s modular and composable design as they could reuse and modify each analysis widget independently. 
However, \system{}'s modular design also introduced friction during initial analysis authoring as it required users to decide at what granularity to specify UI widgets such that they reliably performed their intended behavior and would be reusable across data and tasks. 

We synthesized findings from design iterations into a provisional design framework of three trade-off axes for generative UI in LLM-assisted analysis: \textit{(i) Low vs. high Malleability}---a dimension describing how easy a generative UI can be customized, reconfigured and extended; and \textit{(ii) Implicit vs. Explicit Specification}---a dimension describing how explicit users should specify a generative UI; \textit{(iii) Isolated vs. interoperable}---a dimension describing how a generative UI connects to and exchanges data with others. The contributions of this work are:
\begin{itemize}
    \item Current practice and challenges in using LLM-generated UI for analysis tasks.
    \item \system{}, a technology probe that lets users author and assemble generative widgets on an infinite canvas.
    \item Empirical findings revealing the trade-offs between chat interface and generative UI for analysis tasks.
    \item A three-axis provisional design framework to guide the design of generative UI for LLM-assisted analysis.
\end{itemize}
\section{Related Work}
\subsection{Generative and Malleable UI and LLMs} 
While most LLM products use chat interfaces with text as the primary input and output, LLMs can now generate functioning UI directly from prompts~\cite{chen_generative_2025}, enabling designers and developers to prototype interfaces during early-stage ideation~\cite{lu_misty_2025, chen_genui_2025, noauthor_figma_nodate} using tools like Cursor~\cite{noauthor_cursor_nodate}, v0~\cite{noauthor_v0_nodate}, and Claude Artifacts~\cite{noauthor_claude_nodate}. While direct code generation offers flexibility, generated UIs can be unreliable. Recent work addresses this through specification-driven approaches where LLMs first produce a structured task representation before rendering an interface~\cite{cao_generative_2025, noauthor_introducing_nodate}. Prior generative UI systems provide on-demand adjustment: Potluck turns documents into interactive software~\cite{noauthor_potluck_nodate}, DynaVis synthesizes widgets for iterative visualization editing~\cite{vaithilingam_dynavis_2024}, and Promptions generates UI controls to steer AI responses~\cite{drosos_dynamic_2025}.

Malleable UI research complements generative UI by enabling end-user customization after an interface is generated or deployed. Recent work like Jelly~\cite{cao_generative_2025} proposes frameworks of evolving task-driven data models that let users modify the task data (e.g., plans for a dinner party) either directly through UI or using LLMs. Malleable overview-detail interfaces show how users can customize what appears in overview and detail views and how these views are composed and laid out, rather than being confined into a designer-chosen form~\cite{min_malleable_2025, min_meridian_2025}. 

These systems typically treat the generative UI as an ephemeral solution for a single session (e.g., a prototype, a visualization edit). Our work further probes how to move from ephemeral generative UI to reusable generative UI that can be parameterized and reused across data and sessions. 

\subsection{LLM-Assisted Analysis}
LLMs can support analysis practices like data processing~\cite{shankar_steering_2025} and information seeking~\cite{10.1145/3706598.3714057} across data work~\cite{gu_how_2024, payandeh_noteex_2025, weng_insightlens_2025} and broader knowledge work~\cite{yun_generative_2025, zhang_visar_2023, dellacqua_navigating_2023} in both personal~\cite{ma_proactiveagent_2023} and professional contexts~\cite{noauthor_user_nodate, zhang_rethinking_2024}. Prior HCI systems integrate LLMs into computational notebooks to support data exploration~\cite {payandeh_noteex_2025} and analysis~\cite{freund_flowco_2025, gu_how_2024}. For example, Flowco integrates LLMs into computational notebooks to author end-to-end structured data analysis workflows~\cite{freund_flowco_2025}. 
Besides structured tabular data, most data in working space is unstructured. Recent work like DocWrangler~\cite{shankar_steering_2025} and CollabCoder~\cite{gao_collabcoder_2024} leverages LLMs to semantically process unstructured data and generate insights. 

Prior empirical studies with analysts~\cite{gu_how_2024, kasica_dirty_2023} and product managers~\cite{yun_generative_2025} reveal practices and design considerations for human-AI collaboration in analysis tasks. Data analysts often share and reuse data processing scripts~\cite{zhang_how_2020, muller_how_2019} and synthesize information from diverse sources (e.g., documents, images, structured datasets) that require different analytical tools and approaches~\cite{kasica_dirty_2023}. To support the design of LLM-powered analysis tools, Gu et. al. suggest that they should fluidly connect data and procedure at various granularity and communicate data operations through diverse outputs (\emph{e.g.}, code, natural language descriptions, visualizations)~\cite{gu_how_2024}. Yun et. al. further highlight the necessity to support diverse workflows and individual preferences~\cite{yun_generative_2025}.
LLMs can now generate diverse data analysis operations, enabling users to customize analyses beyond the predefined operations in traditional node-based systems (\emph{e.g.}, map, filter, and lambda). Our work studies the design considerations and trade-offs in supporting user-defined analysis operations.
\begin{table*}[h]
\centering
\resizebox{\textwidth}{!}{%
\begin{tabular}{@{}llllll@{}}
\toprule
\textbf{PID} & \textbf{Age} & \textbf{Gender} & \textbf{Profession} & \textbf{Coding Exp.} & \textbf{Current LLM Use} \\ \midrule
F1 & 25 & F & Program Manager & Yes & Synthesize insights from unstructured data; Search \\
F2 & 26 & M & Business Analyst & Yes & Analyze sales trend; Generate SQL for analysis \\
F3 & 32 & M & Management Consultant & Yes & Synthesize insights from unstructured data; Generate python for analysis \\
F4 & 24 & M & Law Student & No & Distill central claims of past cases \\
F5 & 29 & F & Journalist & No & Cleanup interview transcripts; Outline stories\\
F6 & 33 & F & Middle School Teacher & No & Brainstorm class activities; Create assessments; Summarize feedback \\ \bottomrule
\end{tabular}%
}
\caption{Participant demographics of the formative study, including age, gender, profession, coding experience and current LLM use.}
\label{tab:formative-participant}
\end{table*}


\section{ Current Practice and Usage of Generative UI}
Our goal is to understand the current and potential future use of LLM-generated GUI tools for analysis work. 
We conducted a two-part formative study: \textit{(i)} semi-structured interviews\footnote{The study protocol was approved by the IRB of our institution.} with 6 professionals to understand how people currently use LLM chat assistants and their feedback on creating LLM-generated GUI tools, and
\textit{(ii)} an analysis of publicly available, user-created LLM-generated GUI tools to examine usage contexts and design patterns. We then synthesized design opportunities by reflecting on findings of both studies. 

\begin{figure}[htb]
  \centering
  \includegraphics[width=3.33in]{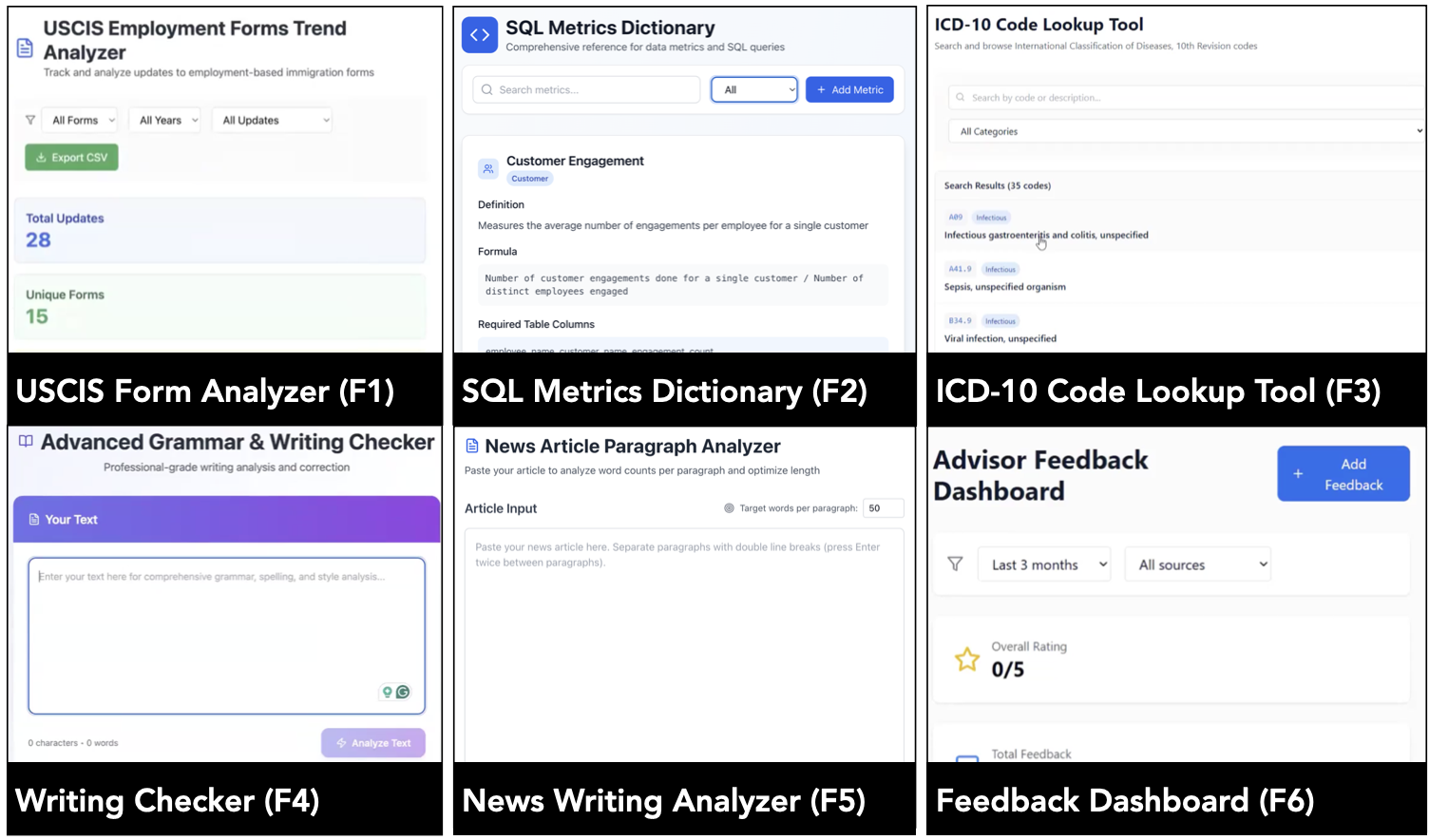}
  \caption{Screen shots and the names of the Claude artifacts made by 6 formative study participants.}
  \label{fig:formative}
\end{figure}

\subsection{Formative Interviews}
We recruited 6 participants (3 Male, 3 Female) from diverse professional backgrounds who often use LLMs to assist their everyday work through Upwork\footnote{\url{https://www.upwork.com/}} and community networks (Table~\ref{tab:formative-participant}). We compensated participants 30 USD for a 60-minute study on Zoom. All participants had prior experience with generative AI tools such as ChatGPT at work. Three of the participants had coding experience for data analysis (F1, F2, F3) (\textit{e.g.}, Python, SQL) and three did not have any prior coding experience. None of the participants had experience with web or user interface development. 
The study consisted of \emph{(i)} a semi-structured interview and \emph{(ii)} a task of LLM-generated GUI creation using Claude Artifacts. We asked participants about use cases and key friction points in using LLM to analyze workplace data. In the artifact creation task, participants used Claude Artifact to create a tool to address an issue in their workflows. The interview concluded with a discussion on the benefits and challenges of using LLM-generated GUIs in assisting their actual work. We recorded and transcribed the interviews and analyzed them via affinity diagramming~\cite{braun_using_2006}. 
\\

\noindent\textbf{\textit{Current practice of using LLMs in day-to-day analysis.}} 
All participants extensively used LLMs for quantitative (F2, F3) and qualitative data (F1-6) analysis. People also used LLMs for search (F1), idea generation (F5, F6), and to complement common productivity tools (Table~\ref{tab:formative-participant}). 
Participants (F1, F3) reported that LLMs reduced frustration when wrangling heterogeneous and unstructured data compared to traditional tools like Tableau or Excel~\cite{kandel_wrangler_2011}. 
All of the participants mentioned that they have common analysis prompts that they frequently reuse, yet \textbf{reusing these analyses in chat windows is challenging}. Four of six participants (F1-3, F5) frequently reused prior analysis by \textbf{rewriting the same prompts from scratch} each time. For example, F2, who uses generative AI to analyze business data, said, \textit{``ChatGPT is not fine-tuned to business. But business has so many customized definitions and metrics. So every time I want to ask for some predefined calculations, I have to explain the definitions and then the calculation formula that derives the final metrics. It costs a lot of time.''} On the other hand, F4 developed a workaround by pinning prior successful chats and reusing them as templates for future analyses. He commonly used the law firm's internal LLM to analyze 40 legal cases on toxic chemical developments, \textit{``I definitely don't have time to read all cases, but using AI is also not easy.''} He used a multi-step workflow: he opened multiple chat windows for different analyses, iteratively refined prompts to \textit{“give AI very specific things to work on,”} then reused the saved chats on new cases and synthesized the outputs into a document.
\\

\noindent\textbf{\textit{LLM-generated GUI Creation.}} We invited the same 6 participants to create an artifact to address a recurring analysis task in their workflows with Claude Artifact (Figure~\ref{fig:formative}). 
We specified that the tool should be something that they want to use in their day-to-day work. Participants created dashboard tools to analyze forms (F1, F6), dictionaries to look up common queries or codes (F2, F3), and tools for writing feedback (F4, F5). 
Five of six participants (F1-4, F6) perceived the interactive and visual presentation of information as the major benefit over their typical use of LLMs. 
Three participants (F2-4) reported they were specifically excited about the interactivity that the LLM-generated GUIs provide. They typed queries or pasted writing samples into the input field and noticed that the interface returned the information (e.g., SQL code) or feedback (e.g., wording critique) that they wanted. 

After using their generated GUIs, the five of six participants (F1-4, F6) who were initially excited about Claude Artifact encountered several issues when attempting to refine their GUIs. The shared problem was that \textbf{LLM-generated GUIs tightly couple data, analysis, and UI together}, creating interfaces that resist adaptation. For example, F2 created a SQL Metrics Dictionary that converts business formulas into SQL code: \textit{``It looks great for the sample data in the artifact. But when I tested with some other formula, I realized it just hard-coded [the examples].''} When he attempted to type new formulas beyond the initial examples, he discovered the tool did not process the input. \textit{"I  just plug it in—I have to start over completely"}  (F3). F1 also noticed that the form stored in the USCIS Form Analyzer only contains 2023 and 2024, while she wanted to analyze the trend of the changes of the USCIS Employment Form over the past decade.
F6, a middle school English teacher, experienced overhead when using the Advisor Feedback Dashboard she created. \textit{``I thought having a user interface would make things easier''}, but the interface requires her to manually log advisor feedback into the system. \textit{``I'm spending time both explaining what I want and manually logging data every time... that will kill me.''} (F6) The tool did not actually simplify her work but reinvented the wheel. This tight coupling between data and functionality means that participants need to \textbf{re-explain their task context and manually input data for each new instance}, paradoxically increasing rather than reducing their workload. 
All participants recognized that current LLM-generated GUIs promised a way to create customized tools to assist their workflows in principle, but encountered significant barriers to using LLM-generated GUIs in practice.
\\

\subsection{Analysis of Existing LLM-Generated GUIs}
We randomly sampled 100 of 934 user-created tools in the ``Data Analysis'' category on Claude Artifacts.\footnote{\url{https://claude.ai/catalog/artifacts/data-analysis} Link no longer active; artifact list provided in Supplementary Material.} Two researchers independently open-coded each tool's \emph{input data}, \emph{output format}, \emph{interactivity}, and \emph{purpose}, then reconciled codes via affinity diagramming. Results are summarized in Table~\ref{tab:artifact_patterns}. Most artifacts handle mixed or unstructured data (91\%) and produce infographics (75\%) that combined quantitative data visualization and insights from the qualitative data, while structured input formats (tables, JSON) are rare (7\%). Only 20\% provide interactive controls beyond basic navigation, and nearly three-quarters (72\%) serve primarily visualization rather than active analysis.

\begin{table}[t]
\centering
\small
\setlength{\tabcolsep}{6pt}
\renewcommand{\arraystretch}{1.08}
\begin{tabular}{llr}
\toprule
\textbf{Dimension} & \textbf{Category} & \textbf{Count} \\
\midrule
\multirow{5}{*}{Input data}
& Mixed qualitative + quantitative data & 68 \\
& Text data & 23 \\
& JSON files & 4 \\
& Tables & 3 \\
& No input data & 2 \\
\midrule
\multirow{5}{*}{Output format}
& Infographic & 75 \\
& Text report & 16 \\
& Multiple quant. visualization & 6 \\
& Single visualization & 3 \\
\midrule
\multirow{4}{*}{Interactivity}
& None & 59 \\
& Navigation controls mainly & 21 \\
& Visualization controls mainly & 16 \\
& Input fields mainly & 4 \\
\midrule
\multirow{3}{*}{Purpose}
& Visualization & 72 \\
& Deriving insights & 24 \\
& Webpage mockup & 4 \\
\bottomrule
\end{tabular}
\caption{Observed characteristics of 100 sampled user-created data-analysis GUI artifacts.}
\label{tab:artifact_patterns}
\end{table}

\subsection{Design Opportunities}
Our formative study with professionals using LLMs and analysis of existing LLM-generated GUIs revealed the following design opportunities for LLM-generated GUIs: \\ 
\begin{figure*}[htb]
  \centering
  \includegraphics[width=0.95\textwidth]{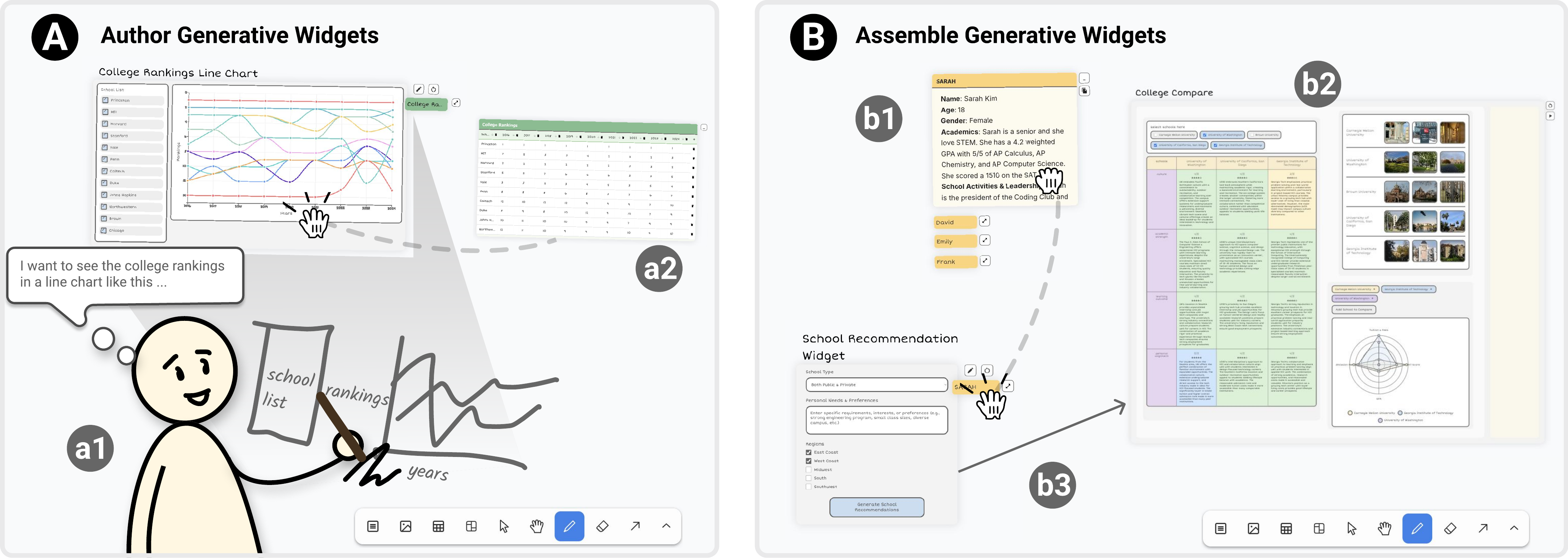}
  \caption{Alice, a college counselor, uses \system{} to support student advising. She sketches and describes to (a1) author a Line Chart widget. She can now (a2) drag college ranking data onto the widget to visualize ranking trajectories. She also previously (b1) created a School Recommendation widget to get personalized college suggestion. As Alice and her student discuss options, she (b2) assembles a Radar Chart, Comparison Table, and Campus Image widget into a container for a side-by-side composite view, then (b3) connects the recommendation widget to the container with an arrow to create a reusable pipeline for future students.}
  \label{fig:interface}
\end{figure*}

\noindent \textit{\textbf{D1: Support heterogeneous and unstructured input data. }} Participants in our formative interviews identified that LLM chatbots provided benefit over traditional analysis workflows, particularly for heterogeneous and unstructured data. Similarly, our artifact analysis revealed that people currently using LLM-generated GUIs also created such customized GUIs most often for mixed data types and second most often for unstructured text data. Thus, systems to support generative UIs for analysis should \textbf{prioritize input flexibility} to handle heterogeneous and unstructured data. \\ 

\noindent \textit{\textbf{D2: Support idiosyncratic analysis needs. }} LLM-generated GUIs created by our formative interview participants and analyzed in our artifact analysis were each unique from one another and not supported by traditional analysis tools. 
Our findings suggest an unmet need for analysis tools that support a long tail of analysis needs. 
Systems to support LLM-generated GUIs should \textbf{preserve expressivity} such that people can create personal analysis tools. \\ 

\noindent \textit{\textbf{D3: Support reuse of prior analysis. }} Our formative interview participants frequently reused analyses but found chatbots poorly supported this, resorting to rewriting prompts or manual workarounds. Both formative interview and publicly-available artifacts revealed the same barrier: GUIs tightly couple input data, analysis logic, and display, making reuse across datasets non-trivial.
Future systems should \textbf{decouple data, analysis, and display} to allow people to easily reuse their analyses. \\ 

\noindent \textit{\textbf{D4: Support modification. }} Our formative interview participants first created LLM-generated GUIs by providing a prompt, but often found the initial GUI did not meet their expectations such that they would then provide editing instructions or rewrite the initial prompt. Future systems to support LLM-generated GUIs should \textbf{encourage modularity} to let users edit one part of the GUI without impacting the entire GUI in unpredictable ways. \\

\noindent Thus, we aimed to create a technology probe that preserves the flexibility (\textbf{D1}) and expressivity (\textbf{D2}) of general-purpose LLM-generated GUI applications, but introduces support for reuse (\textbf{D3}) and modification (\textbf{D4}) that are particularly useful for frequent and multi-step analysis workflows encountered during our formative work.
\section{Technology Probe: \system{}}
We introduce \system{}, a technology probe for \textbf{authoring} and \textbf{assembling} LLM-generated GUI widgets on an infinite canvas. Figure~\ref{fig:interface} and the video in Supplementary Material provide walkthroughs.

\subsection{Data Shapes and Widgets}
\system{}'s design is inspired by and grounded in the instrumental interaction model ~\cite{beaudouin-lafon_instrumental_2000}, which describes GUIs as \emph{domain objects} (content or data of the task) and \emph{interaction instruments} (tools the user manipulates to act on domain objects). Similarly, \system{} defines two kinds of shapes on canvas: \emph{(i)} \textbf{Data Shapes}: the \emph{domain objects} that contain \textbf{data} (\emph{e.g.,} articles, images) of the task, and \emph{(ii)} \textbf{Widget}: user-created \emph{interaction instruments} that carry out specific \textbf{analysis} on data shapes. This separation of analysis from data also follows the principles of instrumental interaction~\cite{beaudouin-lafon_reification_2000}: \emph{(1) reification}--- make individual analysis as interactive widgets; \emph{(2) polymorphism}--- the same widget can operate on multiple data shape types; \emph{(3) reuse}--- the user can reuse a generative widget for similar analysis. Below, we detail our design and components of data shapes and widgets. 

\subsubsection{Data Shapes}
\system{} supports three common data types: text, images, and tabular data. These data types are general enough to capture a wide range of everyday tasks: text for unstructured notes, images for visual information such as photos, diagrams, and webpage screenshots, and tabular data for structured data. Accordingly, we designed \textbf{Text Shape} that supports markdown editing and rendering; \textbf{Image Shape} that displays visual content; and \textbf{CSV Shape} that allows users to view tabular data. Users can freely add, edit, copy, and delete these data shapes on the canvas. Users can select one of the shape creation tools on the toolbar ( \raisebox{-2pt}{\includegraphics[scale=0.19]{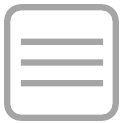}} for text, \raisebox{-2pt}{\includegraphics[scale=0.19]{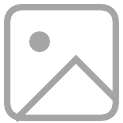}} for image, and \raisebox{-2pt}{\includegraphics[scale=0.19]{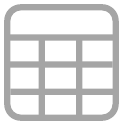}} for CSV) (Figure~\ref{fig:interface}A) and then upload data. 

\subsubsection{Widget}
Widgets are interactive GUI tool that carries out an analysis on data shapes. A widget takes a data shape input, and then provides information seeking(\emph{e.g.,} recommend three relevant papers about this topic), visualization (\emph{e.g.,} visualize an itinerary on a map), analysis (\emph{e.g.,} analyze strengths, weaknesses, opportunities, and threats for a product) or other analysis. Widgets mainly consist of a \textbf{GUI component}, a \textbf{LLM component}, and a \textbf{schema} that connect them together (Figure~\ref{fig:widget-architecture}). See Supplementary Material for
an example architecture of a travel itinerary generative widget.
\\
\begin{figure}[htb]
  \centering
  \includegraphics[width=3.5in]{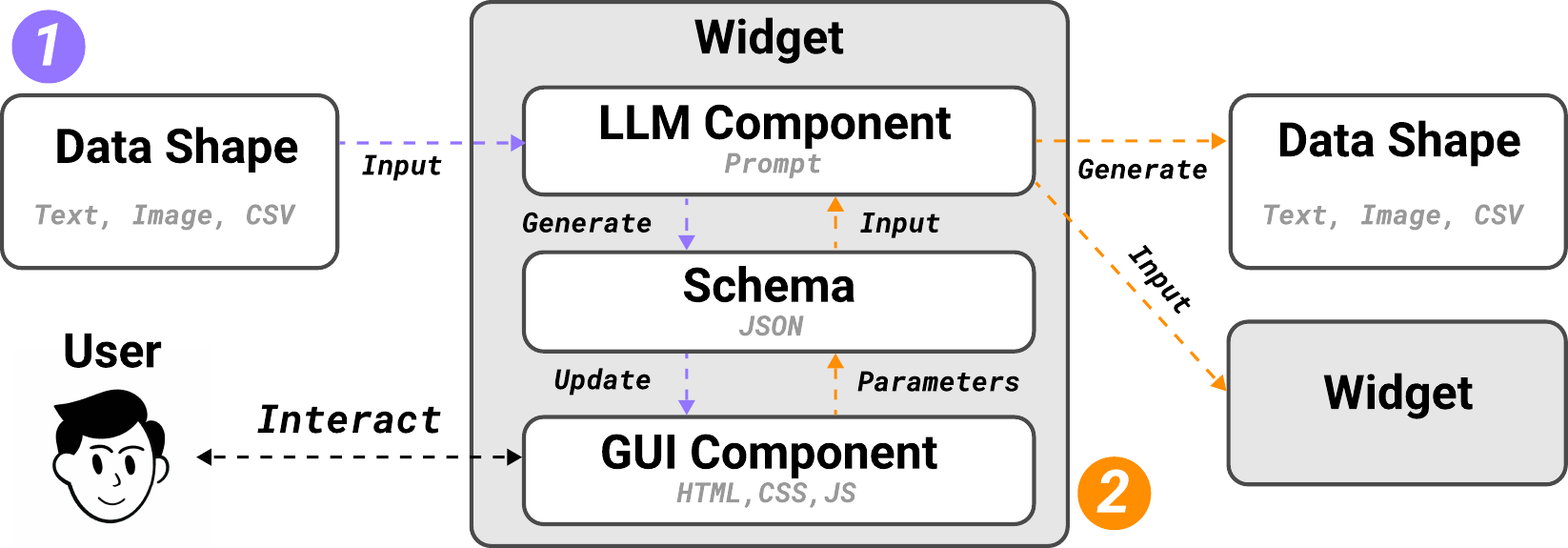}
  \caption{Widget Architecture. A shared schema coordinates the LLM and GUI components in (1) data-first, interaction-second and (2) interaction-first, data-second workflows.}
  \label{fig:widget-architecture}
\end{figure}

\noindent\textbf{GUI Component.} The GUI component provides user controls and defines the visual layout and interaction affordances of the widget. The GUI component supports users in interacting, such as adding new key points to a list of ideas, filtering the radar chart, or zooming in on maps. The visual layout is interpreted from user prompts and sketches, providing contextually appropriate interface elements for the intended analysis. As we focus on functionality rather than styling, we use predefined CSS guidelines to avoid excessive token usage on unnecessary UI design. 

\noindent\textbf{LLM Component.} The LLM component is responsible for the core data processing and transformation logic that determines how the widget operates on incoming data. It takes the provided input data together with the prompt and executes the specified analysis to produce the desired output.

\noindent\textbf{Schema.} The schema enables the communications between the GUI and the LLM components, as it defines \emph{(i)} how LLM-processed data outputs are structured for display in the GUI, and \emph{(ii)} how parameters and context set in the GUI are embedded in prompts. 

\noindent\raisebox{-2pt}{\includegraphics[scale=0.3]{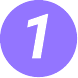}}~\textbf{\textit{Data-First, Interaction Second:}} The LLM component first processes and transforms input data objects, then displays them in new representations through the GUI component (\emph{e.g.,} a map viewer that visualizes an itinerary as routes and waypoints). Widgets first process input and then visualize output, serving primarily as customized displays for complex data, similar to prior work~\cite{cao_generative_2025, min_malleable_2025}. 

\noindent\raisebox{-2pt}{\includegraphics[scale=0.3]{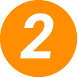}}~\textbf{\textit{Interaction First, Data-Second:}} The GUI component first takes user-defined parameters and context, then the LLM component creates new data shapes or passes the generated data to other widgets(\emph{e.g.,} a story composer where users input key ideas and select tone or style). GUIs serve as control panels for generation, allowing users to guide the prompt's output through structured input, similar to prior work~\cite{vaithilingam_dynavis_2024}.

\subsection{Interface Features and Interactions} 
In this section, we introduce the feature of \system{} by following the steps of authoring, using, assembling, and refining widgets.
\\

\noindent\textbf{\textit{Authoring Widgets.}} Users can use natural language descriptions and sketches to describe the visual and behavioral aspects of the widget (Figure~\ref{fig:interface}A). \system{} also supports voice input so that the user can sketch and talk at the same time to facilitate efficiency and facilitate thinking during the process ~\cite{rosenberg_drawtalking_2024, krosnick_think-aloud_2021}.
\\

\begin{figure}[htb]
  \centering
  \includegraphics[width=3.33in]{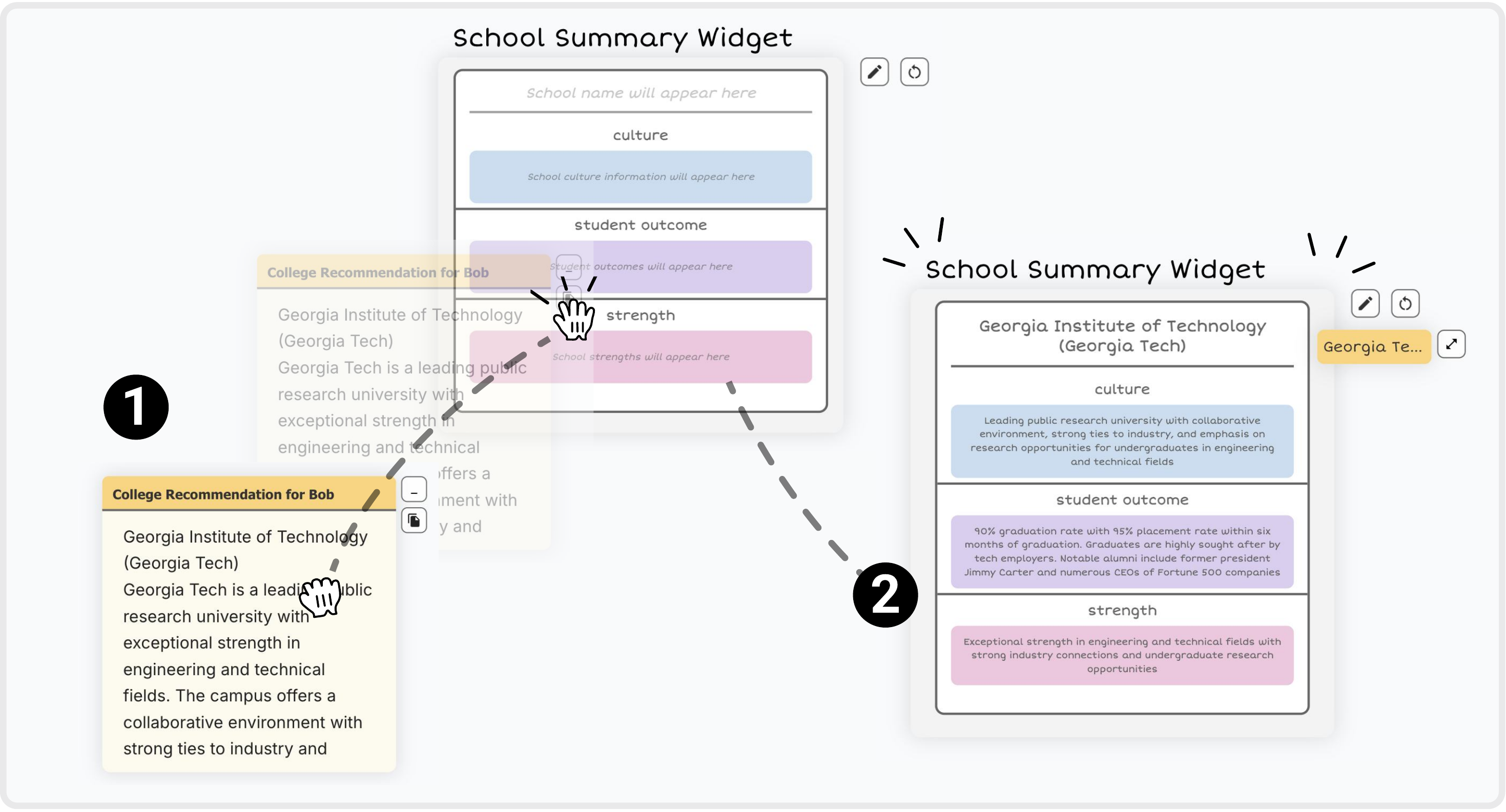}
  \caption{Example of a drag-and-drop interaction. (1) Users drag-and-drop the ``College Recommendation for Bob'' Text Shape to the School Summary Widget. (2) School Summary Widget automatically extracts and summarizes information about Georgia Tech and displays Georgia Tech's culture, student outcomes, and strengths in the corresponding field on the widget.}
  \label{fig:drag-and-drop}
\end{figure}

\begin{figure}[htb]
  \centering
  \includegraphics[height=3.33in]{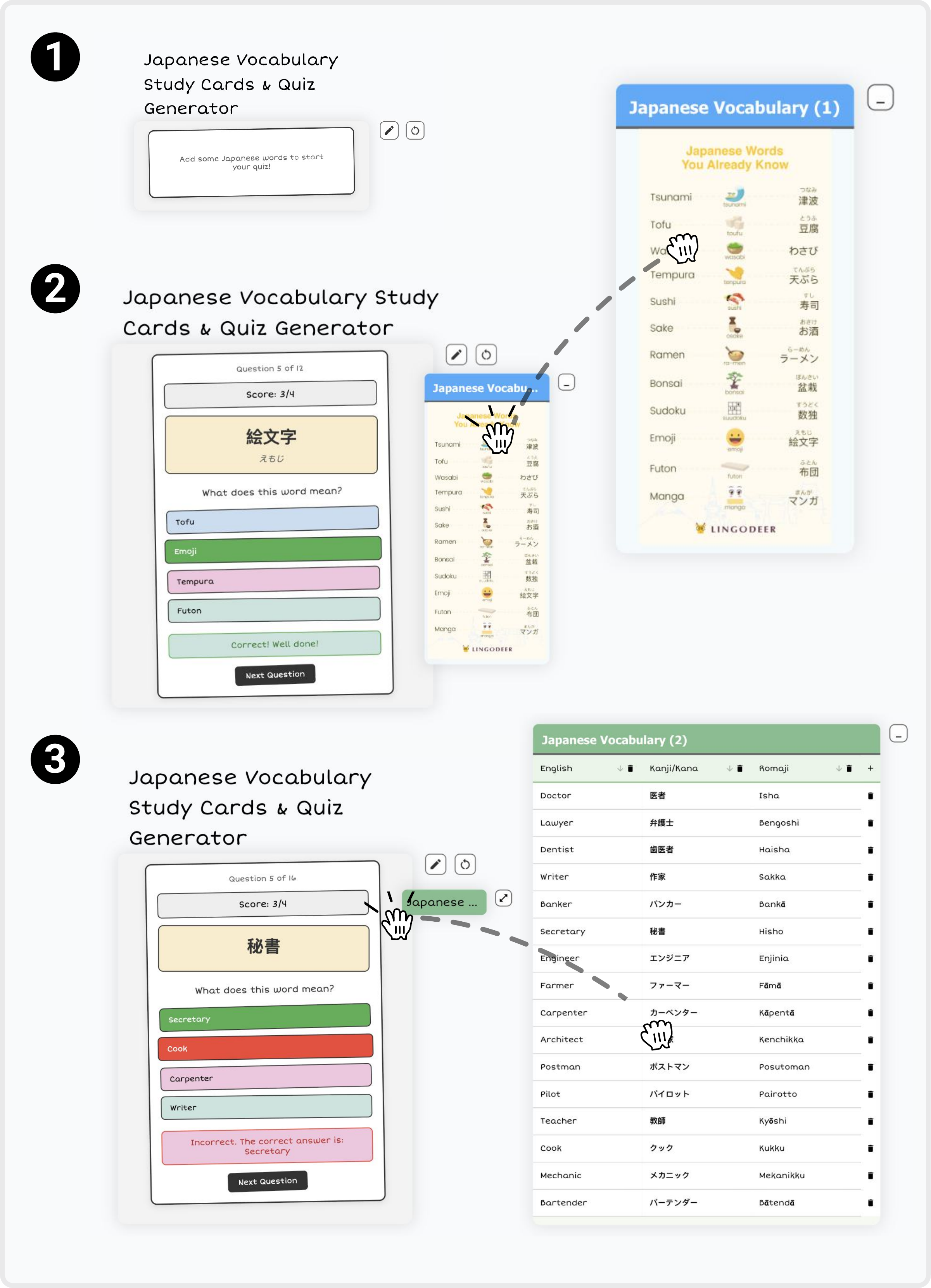}
  \caption{Reuse of ``Japanese Vocabulary Study Cards \& Quiz'' widget with different input data types. (1) The widget is in the idle state with no internal data. (2) The widget generates quiz cards from a Japanese vocabulary list in an \textbf{Image Shape}. (3) The same widget can bind to another list of Japanese words in a \textbf{CSV Shape}.}
  \label{fig:reuse}
\end{figure}

\noindent\textbf{\textit{Using Widgets.}} Users can drag data shapes onto widgets to bind the data to widgets (Figure~\ref{fig:drag-and-drop}) to process data, and then users can interact with it. If the widget has a control panel for parameter input (\emph{e.g.,} dragging a student profile document onto a recommendation widget and selecting preferences), users can adjust parameters, then click the button on the widget to process the data (Figure~\ref{fig:interface} B). Users can also reuse the same widget for other data(Figure~\ref{fig:reuse}). As the generative widget may contain errors, users can click the {\includegraphics[scale=0.19]{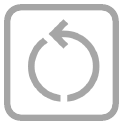}} button to clear all data in the widget and revert it to the idle state.
\\

\noindent\textbf{\textit{Assembling Widgets}} \system{} supports two composition patterns that enable complex workflow construction (Figure~\ref{fig:interface}B): 
\begin{itemize}
    \item \textbf{\textit{Fan-out Composition:}} Users can use \raisebox{-2pt}{\includegraphics[scale=0.19]{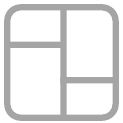}} button on the toolbar (Figure~\ref{fig:interface}B)  to create a container object. It can ``bundle'' to process the same input data simultaneously. For example, bundling a visualization widget and a summary widget to analyze school information from multiple perspectives simultaneously (Figure~\ref{fig:interface} b2).
    \item \textbf{\textit{Sequential Composition:}} Users can use \raisebox{-2pt}{\includegraphics[scale=0.19]{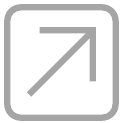}} on the toolbar (Figure~\ref{fig:interface} A) to draw an arrow from one widget to another widget or container. The output of the widget will be automatically passed to another's, creating processing pipelines. Visual connection lines indicate data flow direction, and users can chain multiple widgets together by connecting widgets and containers (Figure~\ref{fig:interface} b3).
\end{itemize}

\begin{figure}[htb]
  \centering
  \includegraphics[height=3.33in]{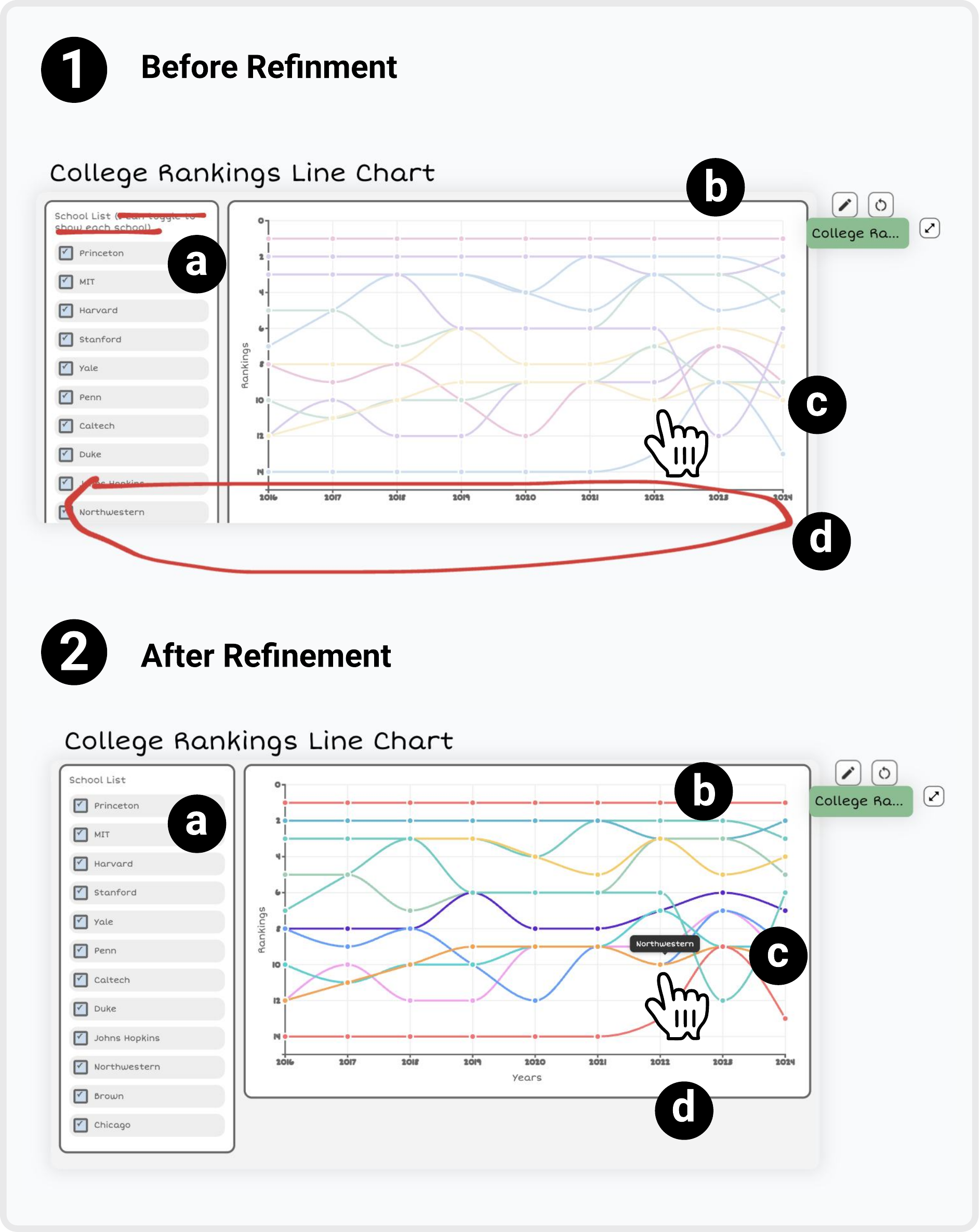}
  \caption{Refinement of the College Rankings Line Chart widget using sketch annotation (red strokes) and description. (1) Before Refinement: (a) Irrelevant text on the title of the school list; (b) line colors are too faint; (c) no tooltip when hovering over a line; (d) overflow of height. (2) After Refinement: (a) Cleaned and expanded school list; (b) stronger, more distinguishable line colors; (c) interactive tooltip shown on hover; (d) properly adjusted height and layout.}
  \label{fig:refine}
\end{figure}

\noindent\textbf{\textit{Refining Widgets.}} When the widget does not meet the user's goal (Figure ~\ref{fig:refine}), or the user wants to repurpose a widget for other similar purposes (\emph{e.g.,} change a Japanese phrase flip card widget to a biology concept flip card widget), they can further use sketch and description to update the widget. The GUI component, LLM component, and schema will all be updated. Users can also directly edit the prompt by clicking \raisebox{-2pt}{\includegraphics[scale=0.19]{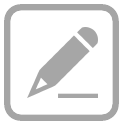}} at the top right corner of the widget (\emph{e.g.,} rewrite the prompt in LLM component from ``summarize this document in 3 paragraphs'' to ``extract only the key findings and format as bullet points'').

\subsection{Implementation}
\label{sec:implementation}

\system{} was implemented as a React and TypeScript web application. The infinite-canvas interface and custom shape classes were built using the TLDraw SDK\footnote{\url{https://www.tldraw.com}}
. Each widget appears on the canvas as a webpage embedded in a \texttt{iframe}. We utilize OpenAI (\texttt{gpt-4o}) to interpret the sketch and language description transcribed from the browser's speech recognition. We use Anthropic’s \texttt{Claude-Sonnet-4.0} for code generation; OpenAI’s \texttt{GPT-4o} with web-search and code-execution tools powers the LLM components in generative widgets. We use the Unsplash API\footnote{\url{https://unsplash.com/}} for image retrieval. 
\section{User Study}
We conducted a within-subject controlled study comparing \system{} to a chat baseline for creating and reusing multi-step analysis workflows~\cite{rule_exploration_2018} with \numpar{} professionals (4F, 8M; ages 23--60) from Upwork\footnote{\url{https://www.upwork.com/}} and community networks. All were knowledge workers spanning business, creative, technology, and research fields who used generative AI tools daily or weekly for data analysis (Table~\ref{tab:user-study-participant}).
We asked:

\begin{itemize}[
    label=\textbf{Step \arabic*},
    leftmargin=*,
    labelsep=0.6em,
    itemsep=0.2em
]
\item [\textbf{RQ1}] What is the performance of workflow creation and reuse for \system{} and the baseline?
\item [\textbf{RQ2}] What are the benefits and limitations of \system{}’s modular, composable generative widget design?
\end{itemize}

\begin{table}[t]
\footnotesize
\setlength{\tabcolsep}{3pt}
\renewcommand{\arraystretch}{1.05}
\begin{tabularx}{\columnwidth}{c c c l c X}
\toprule
\textbf{PID} & \textbf{Age} & \textbf{Gen.} & \textbf{Profession} & \textbf{Code Exp. (yrs)} & \textbf{GenAI Tools} \\
\midrule
P1  & 39 & M & Web Content Writer  & 3 & Ch., G, P, D \\
P2  & 26 & F & PhD Student         & 3 & Ch. \\
P3  & 36 & M & UX Designer         & 2 & Ch. \\
P4  & 40 & F & Business Analyst    & 0 & Ch., Cl. \\
P5  & 23 & M & Game Designer       & 5 & Ch., C, G, D \\
P6  & 30 & F & Mgmt. Consultant    & 0 & Ch., C, G \\
P7  & 28 & M & Project Manager     & 6 & Ch., C, Cl. \\
P8  & 24 & M & PhD Student         & 8 & Ch., C \\
P9  & 60 & F & Design Researcher   & 0 & Ch., C, G \\
P10 & 28 & M & Software Engineer   & 8 & Ch. \\
P11 & 23 & M & Software Engineer   & 5 & Ch., C \\
P12 & 37 & M & Creative Consultant & 1 & Ch. \\
\bottomrule
\end{tabularx}
\caption{Participant demographics for the user study (\(N=\numpar{}\)). GenAI tools are abbreviated as Ch. = ChatGPT, C = Claude, G = Gemini, Cl. = Copilot, P = Perplexity, and D = Deepseek.}
\label{tab:user-study-participant}
\end{table}

\subsection{Study Design}
\begin{figure}[t]
  \centering
  \includegraphics[width=3.33in]{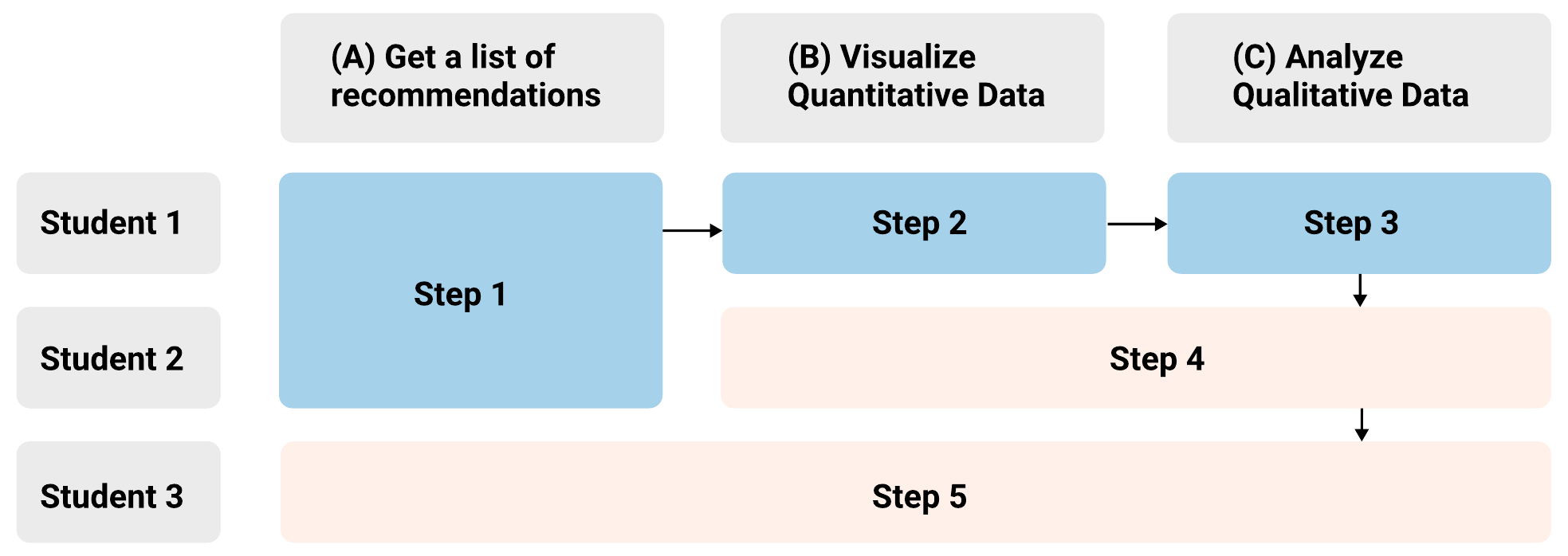}
    \caption{\textbf{User Study Design.} Participants first complete an analysis task (Steps~1--3), then repeat it (Steps~4--5).}
  \label{fig:study-design}
  \vspace{-11pt}
\end{figure}

We designed two realistic decision-making tasks: a career recommendation task (T1) and a college recommendation task (T2), each requiring participants to act as a counselor and recommend options to 3 students by analyzing both quantitative and qualitative data across repeated steps (Figure~\ref{fig:study-design}). 
Steps 1--3 establish the analysis workflow: get recommendations, visualize quantitative data (\emph{e.g.} expected salary, admission rates), and analyze qualitative aspects(\emph{e.g.} required knowledge, school culture) for Student 1; Steps 4--5 repeat this workflow for Students 2 and 3, testing reuse. Student personas were generated using \texttt{GPT-4o}~\cite{chen_why_2025, chen_empathy-based_2024}. Tasks can be found in the Supplementary Material.

We used Claude Artifacts (\texttt{claude-sonnet-4.0}) as baselines by default because it was the state-of-the-art chat-based system that could generate interactive UIs for data analysis by the time the study was conducted.~\footnote{P2 hit the Claude token limit and used ChatGPT Data Analyst, another state-of-the-art chat-based system that could generate interactive data visualizations.} Each participant completed both conditions in counterbalanced order to ensure any differences in performance reflect interaction designs of \system{} rather than models or study procedure. Participants imported materials and authored widgets from scratch, with no pre-loaded content.

\subsection{Procedure and Analysis}
Sessions lasted 1.5 hours over Zoom (IRB approved; \$45/hr compensation). Each began with a pre-study questionnaire and a 5-minute \system{} tutorial, followed by 50 minutes of counterbalanced tasks. After each task, participants completed a NASA-TLX survey~\cite{hart2006nasa}. Sessions concluded with 10 minutes of free exploration and a semi-structured interview about trade-offs between \system{} and chat interfaces. Qualitative data were analyzed via affinity diagramming~\cite{braun_using_2006}; quantitative comparisons used Wilcoxon signed-rank tests.
\section{Results}
Our results suggest a trade-off: \system{} shifts effort from repeated prompting to upfront workflow authoring. Participants spent more time constructing the initial analysis workflow with \system{}, but less time reusing that workflow for repeated analysis steps.
In this section, we share results of the \textit{(i)} task performance, \textit{(ii)} benefits and challenges introduced by \system{} compared to baseline.

\subsection{Performance Results}
\begin{figure}[h!]
    \centering
    \includegraphics[width=3.33in]{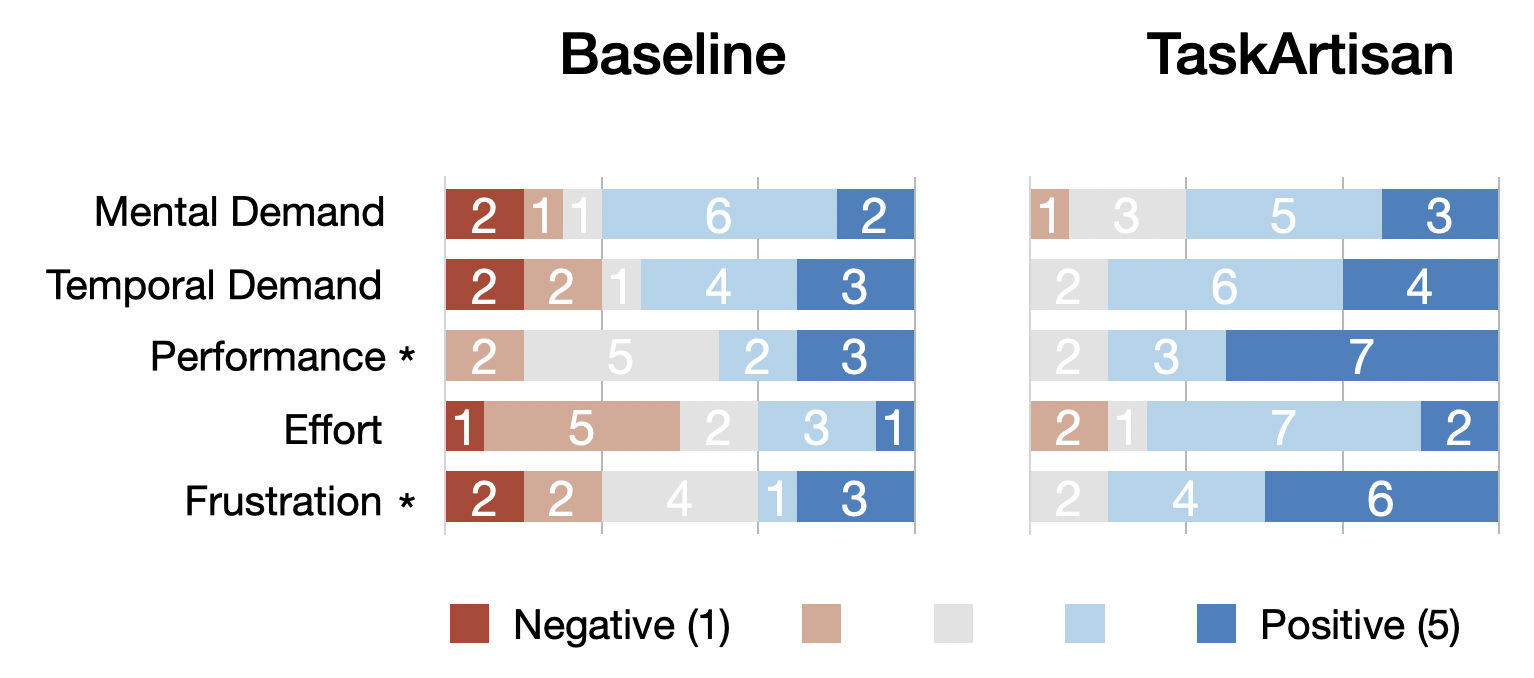}
    \caption{Distribution of the NASA-TLX rating scores for the Baseline and
\system{} (1 = negative, 5 = positive). Statistical significance is marked as * $p < 0.05$ with a Wilcoxon signed-rank test.}
    \label{fig:survey-result}
\end{figure}

10 out of 12 participants using \system{} completed all steps of the tasks with required information, visualization, and analysis, compared to 7 out of 12 participants in the baseline condition (Table ~\ref{tab:step_outcomes_summary}). Participants using \system{} spent significantly \textbf{less time on repetitive steps} (Steps 4--5) than with the baseline (M = 4.3, SD = 2.0 vs. M = 7.8, SD = 2.9 minutes; $p < 0.01$), while spending significantly \textbf{more time setting up analysis steps} (Steps 1--3) (M = 21.1, SD = 4.8 vs. M = 14.0, SD = 3.6 minutes; $p < 0.01$). We observed no significant difference in overall task completion time between \system{} (M = 25.3, SD = 5.3 minutes) and the baseline (M = 21.9, SD = 4.9 minutes). Participants using \system{} reported significantly \textbf{higher performance} (M = 4.41, SD = 0.79 vs. M = 3.5, SD = 1.09; $p < 0.05$) with \textbf{lower frustration}  (M = 1.67, SD = 0.78 vs. M = 2.92, SD = 1.44; $p < 0.05$) in accomplishing our analysis tasks compared to chat-based interface (Figure ~\ref{fig:survey-result}). Participants in the baseline encountered errors from context loss and ambiguous references (e.g., 5/12 failed to get correct visualizations at Step 2), while \system{} users faced reliability issues in widget generation (2/12 incomplete).

\subsection{Benefits and Challenges Introduced by \system{}}
Participants found that GUI-based interaction improved clarity and made analysis steps easier to understand, but the GUI is not as easy to edit as text and requires clear thinking during prompting. Modularity reduces repetitive prompting and makes analysis steps persistent. Yet, it also requires careful abstraction design (e.g., templates, widget libraries, and management tools) to balance reuse with task-specific control. During free exploration, 7 participants created widgets for everyday tasks, while 4 extended workflows from the controlled session. Participants found \system{} particularly useful for iterative analysis and high-dimensional data exploration. See the Supplementary Material for details.
\\

\begin{table}[t]
\centering
\scriptsize
\setlength{\tabcolsep}{11pt}
\renewcommand{\arraystretch}{1.05}
\begin{tabular}{@{}lcccc@{}}
\toprule
Condition & Comp. & Setup$^{**}$ & Repeat$^{**}$ & Total \\
& (\# / 12) & M (SD) & M (SD) & M (SD) \\
\midrule
Baseline  & 7/12  & 14.0 (3.6) & 7.8 (2.9) & 21.9 (4.9) \\
\system{} & 10/12 & 21.1 (4.8) & 4.3 (2.0) & 25.3 (5.3) \\
\bottomrule
\end{tabular}
\caption{Study performance summary. Setup time aggregates Steps 1--3; repeat time aggregates Steps 4--5. All times are in minutes. Statistical significance is marked as $^{**}p < 0.01$ using a Wilcoxon signed-rank test.}
\label{tab:step_outcomes_summary}
\end{table}

\begin{figure}[h!]
    \centering
    \includegraphics[width=3.33in]{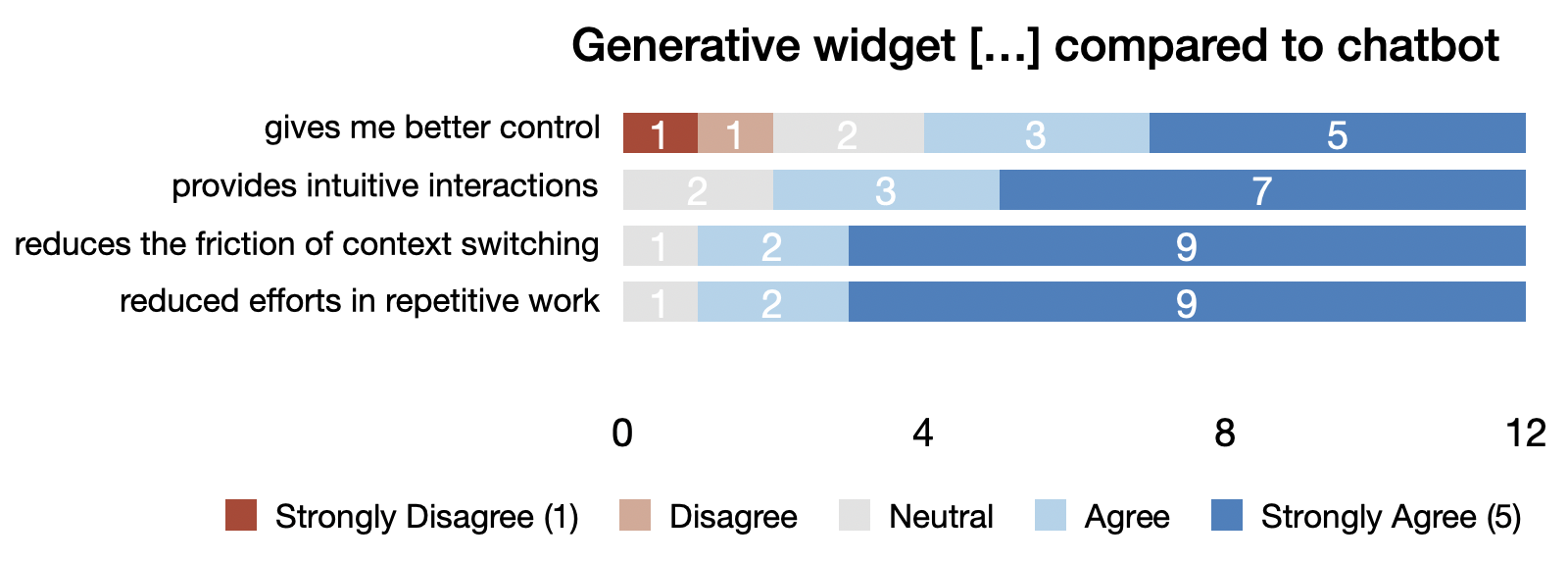}
    \caption{Subjective participant ratings of the usefulness of interactions and system on a scale from 1 (Strongly Disagree) to 5 (Strongly Agree).}
    \label{fig:interaction-result}
\end{figure}

\noindent\textbf{\textit{GUI makes analysis steps clear and visual.}} Participants (10/12) found generative widgets are easier and more intuitive to interact with compared to the chatbot (7 Strongly Agree, 3 Agree, 2 Neutral) (Figure~\ref{fig:interaction-result}). Seven participants (P1, P2, P4, P6, P7, P8, P12) stated that generative widgets make the mapping between data input and analysis results clear. \textit{``When I click the generate school recommendation'' button, I kind of know what I will get''} (P7). Four participants (P2, P4, P6, P7) appreciated that everything is visual in widgets and found \textit{``output easier to understand''} (P2). For instance, P4 reported that displaying jobs' pros and cons on a GUI table makes it easier to interpret and navigate outputs compared to bullet points in chat interfaces. P6 recalled a similar experience of using ChatGPT for trip planning, as it only produces pure text and she could see how \system{} can make the planning more visual. P7 contrasted the experiences: \textit{``ChatGPT can generate thousands of words and requires a lot of reading... On \system{}, you can see all the information in a very intuitive way''} using different visualization widgets.
\\

\noindent\textbf{\textit{GUI are not as easy to edit as text.}} Seven participants (P1, P2, P4, P6, P7, P8, P11) mentioned that generative widgets are not necessary for quick one-off analysis. P8 said GUI is \textit{``too rigid''} for quick information seeking. \textit{``[If I] just want to know the answer or collect information, a chatbot can be useful for that.''} (P8). P4 described the distinction between \system{} and ChatGPT as \textit{``doing visualization Tableau vs. using Python— the product can be seen in a more intuitive way, but not easy to edit''}. Beyond the analysis itself, generative widgets introduce additional uncertainty around visual layout and whether the analysis results will fit and render properly. As a result, P1 and P11 reported a reduced sense of control over widget behavior compared to chatbots (Figure~\ref{fig:interaction-result}). For example, at Step 3, P11 encountered repeated errors when generating a career comparison table: the widget failed to parse the input list of careers into the expected format despite several prompt refinements. Ultimately, he deleted the widget and re-created a new one instead. \textit{``If a widget is generated, it is kind of final. I cannot easily change it.''} (P11). 
\\

\noindent\textbf{\textit{GUI creation requires clear thinking during prompting.}} Creating and refining generative widgets requires thinking, or effort for non-programmers, as it requires people to think about both analysis and UI specification to get a desired result. While \system{} provides sketch and prompt to make authoring easier, participants without programming backgrounds (P4 and P9) found them need to \textit{``think like a software engineer''}(P4). As P4  did not fully specify her request for the radar chart, the drop-down menus to select schools in the radar chart missed a ``select all'' option. She had to manually add all schools to the radar chart she created when comparing them together.
\\

\noindent\textbf{\textit{Modularity makes analysis steps persistent and reliable.}} 
Six participants (P2, P5, P7, P8, P10, P12) highlighted \system{} provided persistent analysis steps compared to chatbots. P3 emphasized that \textit{``when I created a career recommendation widget, it never changes. I can use it from day 1 to day 100.''}  As mentioned before, P2 had this problem when using ChatGPT to complete Step 5 (make school recommendations for the third student): \textit{"When I tried to give recommendations for the third student, GPT started to generate Python code instead of directly displaying [the pie chart]. I had to start a new window to write prompts and complete all tasks again.''} This inconsistency forced her to rebuild her entire analysis workflow. While using \system{}, she appreciated that \textit{``once the widgets are created, you can easily use them for other typical tasks''} (P2). Widgets can \textit{``reliability''} maintain the same flow for different contexts without \textit{``worrying about AI generating random results''}(P6). 
\\

\noindent\textbf{\textit{Modularity reduced repetitive work.}} Most participants (11/12) reported that \system{} reduced repetitive work (9 Strongly Agree, 2 Agree, 1 Neutral) compared to chatbots (Figure~\ref{fig:interaction-result}). Four participants (P1, P2, P3, P6) highlighted how generative widgets eliminate the hurdle to recreate similar prompts from scratch, a common frustration with LLM chatbots. P6 said: \textit{``Do [the same analysis] for another student, I have to do that again with ChatGPT. With \system{}, I only create once and drag the data.''} P1 explained, \textit{``I do not need to copy and paste and carefully craft the instructions again. With ChatGPT, too much time is spent on identifying the problem in the prompt engineering. Now I just drag and drop to reproduce the whole workflow.''}. 
\\

\noindent\textbf{\textit{Modularity requires careful abstraction design.}} Four participants (P4, P5, P10, P12) suggested creating libraries and templates that enable personal workflow and team collaboration. P4 proposed  that the widget templates should allow users to \textit{`customize some part of it.''} Similarly, P10 wanted \textit{`some predefined commonly-used widgets,''} noting that \textit{`we want to be creative but we cannot avoid some basic structures to display the information like tables or pie charts.''} Two participants (P7, P10) also suggested better widget management, such as a sidebar that lists available widgets created by a team and documents describing their expected input data formats and use. At the same time, template design must be handled carefully. P12 highlighted how the level of abstraction determines the control and granularity of analysis that the resulting widget does. \textit{``You can give a very generic input and get a generic widget. You can also give a very task-specific prompt and get a very task-specific widget.''} (P12)
\\

\section{Provisional Design Framework for Generative UI in LLM-Assisted Analysis}

Based on our findings, we propose a three-axis provisional design framework to inform future generative UI design in LLM-assisted analysis (Figure~\ref{fig:framework}). Each axis reflects a trade-off observed in our study.
\\

\noindent\textbf{\textit{Low vs. High Malleability.}} The malleability dimension describes how easily a generative UI can be customized and extended after creation, ranging from a fixed artifact to one supporting user-defined reshaping of both UI and analysis logic. Low-malleability UIs suit stable, well-defined tasks, but create friction when analysis needs evolve mid-workflow, which our participants frequently encountered. At the malleable end, users can adjust analysis granularity, modify UI structure, and recompose workflow steps, but our study found this increases authoring overhead: non-programmers (P4, P9) struggled to specify widgets at the right granularity and needed to ``think like a software engineer.'' Future systems should let users define fixed and customizable parts of UIs~\cite{min_meridian_2025} or enable low-friction manipulations via natural language~\cite{tang_naturaledit_2026}.

\noindent\textbf{\textit{Implicit vs. Explicit Specification.}} The specification dimension describes how much detail users must provide to shape a generative UI, ranging from high-level intent (implicit) to concrete UI layout and analysis logic (explicit). Implicit specification minimizes upfront effort but hides assumptions, leading to mismatches; explicit specification yields predictability but burdens non-programmers. While explicit sketching and prompting helped most participants author functional widgets, those without programming backgrounds found the specification demands unexpectedly high. Neither end of this axis is ideal — future systems could support a middle ground through shared human-LLM representations~\cite{cao_generative_2025, drosos_dynamic_2025, ma_ambigchat_2025} that surface ambiguities incrementally~\cite{leung_squire_2025} rather than requiring complete specification upfront.

\noindent\textbf{\textit{Isolated vs. Interoperable.}} The interoperability dimension describes how a generative UI connects to and exchanges data with other generative UIs rather than functioning as a standalone artifact~\cite{noauthor_graffiti_nodate}. A low-interoperable UI (e.g., Claude Artifacts) takes one input, runs an internal analysis, and presents results only within its own interface — simple and safe, but hard to integrate into real workflows where people move between documents, dashboards, and reporting tools. A high-interoperable UI provides well-defined interfaces to share data across tools, though full interoperability increases engineering complexity and may introduce friction for casual use. Our study showed that participants valued the ability to combine widgets into pipelines (9/12 Strongly Agree), but also encountered robustness failures when LLM-based schema transformation dropped or misformatted items. This suggests interoperability introduces both workflow benefits and reliability costs that future systems must balance — for example, through typed inputs and outputs or \textbf{reactive} composition~\cite{suh_storyensemble_2025} where downstream widgets update automatically.

\section{Conclusion and Future Work}
In this work, we explored design opportunities for making generative UIs useful in analysis workflows. Informed by formative interviews with 6 professionals and an analysis of 100 publicly shared artifacts, we developed \system{}, a technology probe enabling authoring and assembling of interoperable generative widgets. Our user study with \numpar{} participants shows that GUI-based interaction improved clarity and reduced repetitive prompting, but introduced rigidity and higher upfront authoring effort. Modularity helped participants reuse workflows across data, yet required a level of abstraction thinking that non-programmers found demanding. Future work could computationally model and optimize user preferences for each output format (\textit{e.g.}, text, GUI).

We synthesized the observed tensions into three axes (\textit{malleability, specification, interoperability}) as a provisional framework to guide future generative UI design in LLM-assisted analysis. Future research should evaluate generative UI in longer-term, realistic deployments, develop scaffolds for widget management and template reuse, and explore adaptive specification mechanisms that reduce upfront authoring burden. Together, this work takes a step toward understanding how generative UI can move beyond ephemeral chat outputs to reusable, composable analysis tools.




\par\smallskip
\noindent\makebox[\linewidth]{\rule{0.4in}{0.4pt}}
\vspace{-1em}

{\footnotesize
\noindent See the companion Supplementary Material for a video demo, key prompts, and further details of study design and results.\par
}

\section*{Acknowledgment}
\noindent This research was supported by a Google ML and Systems Junior Faculty Award. We also thank our study participants for their time and valuable contribution to this work.

\bibliographystyle{IEEEtran}
\bibliography{references}

@String{Computing = "Computing" }

@String{Computer = "{IEEE} Computer" }

@article{zhang_how_2020,
    title = {How do {Data} {Science} {Workers} {Collaborate}? {Roles}, {Workflows}, and {Tools}},
    volume = {4},
    copyright = {https://www.acm.org/publications/policies/copyright\_policy\#Background},
    issn = {2573-0142},
    shorttitle = {How do {Data} {Science} {Workers} {Collaborate}?},
    url = {https://dl.acm.org/doi/10.1145/3392826},
    doi = {10.1145/3392826},
    language = {en},
    number = {CSCW1},
    urldate = {2025-07-08},
    journal = {Proceedings of the ACM on Human-Computer Interaction},
    author = {Zhang, Amy X. and Muller, Michael and Wang, Dakuo},
    month = may,
    year = {2020},
    note = {Publisher: Association for Computing Machinery (ACM)},
    pages = {1--23},
}

@inproceedings{cao_generative_2025,
    address = {Yokohama Japan},
    title = {Generative and {Malleable} {User} {Interfaces} with {Generative} and {Evolving} {Task}-{Driven} {Data} {Model}},
    copyright = {https://creativecommons.org/licenses/by-nc-sa/4.0/},
    url = {https://dl.acm.org/doi/10.1145/3706598.3713285},
    doi = {10.1145/3706598.3713285},
    language = {en},
    urldate = {2025-07-09},
    booktitle = {Proceedings of the 2025 {CHI} {Conference} on {Human} {Factors} in {Computing} {Systems}},
    publisher = {ACM},
    author = {Cao, Yining and Jiang, Peiling and Xia, Haijun},
    month = apr,
    year = {2025},
    pages = {1--20},
}

@inproceedings{vaithilingam_dynavis_2024,
    address = {Honolulu HI USA},
    title = {{DynaVis}: {Dynamically} {Synthesized} {UI} {Widgets} for {Visualization} {Editing}},
    copyright = {https://www.acm.org/publications/policies/copyright\_policy\#Background},
    shorttitle = {{DynaVis}},
    url = {https://dl.acm.org/doi/10.1145/3613904.3642639},
    doi = {10.1145/3613904.3642639},
    urldate = {2025-07-09},
    booktitle = {Proceedings of the {CHI} {Conference} on {Human} {Factors} in {Computing} {Systems}},
    publisher = {ACM},
    author = {Vaithilingam, Priyan and Glassman, Elena L. and Inala, Jeevana Priya and Wang, Chenglong},
    month = may,
    year = {2024},
    pages = {1--17},
}

@inproceedings{ai_instrument,
author = {Riche, Nathalie and Offenwanger, Anna and Gmeiner, Frederic and Brown, David and Romat, Hugo and Pahud, Michel and Marquardt, Nicolai and Inkpen, Kori and Hinckley, Ken},
title = {AI-Instruments: Embodying Prompts as Instruments to Abstract \& Reflect Graphical Interface Commands as General-Purpose Tools},
year = {2025},
isbn = {9798400713941},
publisher = {Association for Computing Machinery},
address = {New York, NY, USA},
url = {https://doi.org/10.1145/3706598.3714259},
doi = {10.1145/3706598.3714259},
booktitle = {Proceedings of the 2025 CHI Conference on Human Factors in Computing Systems},
articleno = {1104},
numpages = {18},
location = {
},
series = {CHI '25}
}

@inproceedings{rosenberg_drawtalking_2024,
	location = {New York, {NY}, {USA}},
	title = {{DrawTalking}: Building Interactive Worlds by Sketching and Speaking},
	isbn = {979-8-4007-0628-8},
	url = {https://dl.acm.org/doi/10.1145/3654777.3676334},
	doi = {10.1145/3654777.3676334},
	series = {{UIST} '24},
	shorttitle = {{DrawTalking}},
	pages = {1--25},
	booktitle = {Proceedings of the 37th Annual {ACM} Symposium on User Interface Software and Technology},
	publisher = {Association for Computing Machinery},
	author = {Rosenberg, Karl Toby and Kazi, Rubaiat Habib and Wei, Li-Yi and Xia, Haijun and Perlin, Ken},
	urldate = {2025-05-19},
	date = {2024-10-11},
}

@inproceedings{suh_sensecape_2023,
	location = {San Francisco {CA} {USA}},
	title = {Sensecape: Enabling Multilevel Exploration and Sensemaking with Large Language Models},
	isbn = {979-8-4007-0132-0},
	url = {https://dl.acm.org/doi/10.1145/3586183.3606756},
	doi = {10.1145/3586183.3606756},
	shorttitle = {Sensecape},
	eventtitle = {{UIST} '23: The 36th Annual {ACM} Symposium on User Interface Software and Technology},
	pages = {1--18},
	booktitle = {Proceedings of the 36th Annual {ACM} Symposium on User Interface Software and Technology},
	publisher = {{ACM}},
	author = {Suh, Sangho and Min, Bryan and Palani, Srishti and Xia, Haijun},
	urldate = {2025-05-27},
	date = {2023-10-29},
	langid = {english},
}

@inproceedings{beaudouin-lafon_reification_2000,
	location = {Palermo Italy},
	title = {Reification, polymorphism and reuse: three principles for designing visual interfaces},
	isbn = {978-1-58113-252-6},
	url = {https://dl.acm.org/doi/10.1145/345513.345267},
	doi = {10.1145/345513.345267},
	shorttitle = {Reification, polymorphism and reuse},
	eventtitle = {{AVI}00: Advanced Visual Interfaces},
	pages = {102--109},
	booktitle = {Proceedings of the working conference on Advanced visual interfaces},
	publisher = {{ACM}},
	author = {Beaudouin-Lafon, Michel and Mackay, Wendy E.},
	urldate = {2025-05-29},
	date = {2000-05},
	langid = {english},
}

@online{noauthor_cells_nodate,
	title = {Cells, Generators, and Lenses: Design Framework for Object-Oriented Interaction with Large Language Models {\textbar} Proceedings of the 36th Annual {ACM} Symposium on User Interface Software and Technology},
	url = {https://dl.acm.org/doi/10.1145/3586183.3606833},
	urldate = {2025-06-08},
}

@inproceedings{hutchinson_technology_2003,
	location = {New York, {NY}, {USA}},
	title = {Technology probes: inspiring design for and with families},
	isbn = {978-1-58113-630-2},
	url = {https://dl.acm.org/doi/10.1145/642611.642616},
	doi = {10.1145/642611.642616},
	series = {{CHI} '03},
	shorttitle = {Technology probes},
	pages = {17--24},
	booktitle = {Proceedings of the {SIGCHI} Conference on Human Factors in Computing Systems},
	publisher = {Association for Computing Machinery},
	author = {Hutchinson, Hilary and Mackay, Wendy and Westerlund, Bo and Bederson, Benjamin B. and Druin, Allison and Plaisant, Catherine and Beaudouin-Lafon, Michel and Conversy, Stéphane and Evans, Helen and Hansen, Heiko and Roussel, Nicolas and Eiderbäck, Björn},
	urldate = {2025-07-03},
	date = {2003-04-05},
}

@inproceedings{chen_genui_2025,
	location = {Madeira Portugal},
	title = {The {GenUI} Study: Exploring the Design of Generative {UI} Tools to Support {UX} Practitioners and Beyond},
	url = {https://dl.acm.org/doi/10.1145/3715336.3735780},
	doi = {10.1145/3715336.3735780},
	shorttitle = {The {GenUI} Study},
	eventtitle = {{DIS} '25: Designing Interactive Systems Conference},
	pages = {1179--1196},
	booktitle = {Proceedings of the 2025 {ACM} Designing Interactive Systems Conference},
	publisher = {{ACM}},
	author = {Chen, Xiang 'Anthony and Knearem, Tiffany and Li, Yang},
	urldate = {2025-07-07},
	date = {2025-07-05},
}

@inproceedings{muller_how_2019,
	location = {New York, {NY}, {USA}},
	title = {How Data Science Workers Work with Data: Discovery, Capture, Curation, Design, Creation},
	isbn = {978-1-4503-5970-2},
	url = {https://dl.acm.org/doi/10.1145/3290605.3300356},
	doi = {10.1145/3290605.3300356},
	series = {{CHI} '19},
	shorttitle = {How Data Science Workers Work with Data},
	pages = {1--15},
	booktitle = {Proceedings of the 2019 {CHI} Conference on Human Factors in Computing Systems},
	publisher = {Association for Computing Machinery},
	author = {Muller, Michael and Lange, Ingrid and Wang, Dakuo and Piorkowski, David and Tsay, Jason and Liao, Q. Vera and Dugan, Casey and Erickson, Thomas},
	urldate = {2025-07-08},
	date = {2019-05-02},
}

@inproceedings{lu_misty_2025,
	location = {New York, {NY}, {USA}},
	title = {Misty: {UI} Prototyping Through Interactive Conceptual Blending},
	isbn = {979-8-4007-1394-1},
	url = {https://dl.acm.org/doi/10.1145/3706598.3713924},
	doi = {10.1145/3706598.3713924},
	series = {{CHI} '25},
	shorttitle = {Misty},
	pages = {1--17},
	booktitle = {Proceedings of the 2025 {CHI} Conference on Human Factors in Computing Systems},
	publisher = {Association for Computing Machinery},
	author = {Lu, Yuwen and Leung, Alan and Swearngin, Amanda and Nichols, Jeffrey and Barik, Titus},
	urldate = {2025-07-10},
	date = {2025-04-25},
}

@inproceedings{10.1145/3706598.3714057,
author = {Pu, Kevin and Feng, K. J. Kevin and Grossman, Tovi and Hope, Tom and Dalvi Mishra, Bhavana and Latzke, Matt and Bragg, Jonathan and Chang, Joseph Chee and Siangliulue, Pao},
title = {IdeaSynth: Iterative Research Idea Development Through Evolving and Composing Idea Facets with Literature-Grounded Feedback},
year = {2025},
isbn = {9798400713941},
publisher = {Association for Computing Machinery},
address = {New York, NY, USA},
url = {https://doi.org/10.1145/3706598.3714057},
doi = {10.1145/3706598.3714057},
booktitle = {Proceedings of the 2025 CHI Conference on Human Factors in Computing Systems},
articleno = {145},
numpages = {31},
location = {
},
series = {CHI '25}
}

@inproceedings{hart2006nasa,
  title={NASA-task load index (NASA-TLX); 20 years later},
  author={Hart, Sandra G},
  booktitle={Proceedings of the human factors and ergonomics society annual meeting},
  volume={50},
  number={9},
  pages={904--908},
  year={2006},
  organization={Sage publications Sage CA: Los Angeles, CA}
}

@inproceedings{zhang_visar_2023,
    address = {New York, NY, USA},
    series = {{UIST} '23},
    title = {{VISAR}: {A} {Human}-{AI} {Argumentative} {Writing} {Assistant} with {Visual} {Programming} and {Rapid} {Draft} {Prototyping}},
    isbn = {979-8-4007-0132-0},
    shorttitle = {{VISAR}},
    url = {https://dl.acm.org/doi/10.1145/3586183.3606800},
    doi = {10.1145/3586183.3606800},
    urldate = {2025-07-11},
    booktitle = {Proceedings of the 36th {Annual} {ACM} {Symposium} on {User} {Interface} {Software} and {Technology}},
    publisher = {Association for Computing Machinery},
    author = {Zhang, Zheng and Gao, Jie and Dhaliwal, Ranjodh Singh and Li, Toby Jia-Jun},
    month = oct,
    year = {2023},
    pages = {1--30},
}

@inproceedings{beaudouin-lafon_instrumental_2000,
    address = {The Hague The Netherlands},
    title = {Instrumental interaction: an interaction model for designing post-{WIMP} user interfaces},
    copyright = {https://www.acm.org/publications/policies/copyright\_policy\#Background},
    shorttitle = {Instrumental interaction},
    url = {https://dl.acm.org/doi/10.1145/332040.332473},
    doi = {10.1145/332040.332473},
    urldate = {2025-07-20},
    booktitle = {Proceedings of the {SIGCHI} conference on {Human} {Factors} in {Computing} {Systems}},
    publisher = {ACM},
    author = {Beaudouin-Lafon, Michel},
    month = apr,
    year = {2000},
    pages = {446--453},
}

@inproceedings{min_malleable_2025,
    address = {New York, NY, USA},
    series = {{CHI} '25},
    title = {Malleable {Overview}-{Detail} {Interfaces}},
    isbn = {979-8-4007-1394-1},
    url = {https://dl.acm.org/doi/10.1145/3706598.3714164},
    doi = {10.1145/3706598.3714164},
    urldate = {2025-07-22},
    booktitle = {Proceedings of the 2025 {CHI} {Conference} on {Human} {Factors} in {Computing} {Systems}},
    publisher = {Association for Computing Machinery},
    author = {Min, Bryan and Chen, Allen and Cao, Yining and Xia, Haijun},
    month = apr,
    year = {2025},
    pages = {1--25},
}

@inproceedings{krosnick_think-aloud_2021,
    address = {New York, NY, USA},
    series = {{CHI} '21},
    title = {Think-{Aloud} {Computing}: {Supporting} {Rich} and {Low}-{Effort} {Knowledge} {Capture}},
    isbn = {978-1-4503-8096-6},
    shorttitle = {Think-{Aloud} {Computing}},
    url = {https://dl.acm.org/doi/10.1145/3411764.3445066},
    doi = {10.1145/3411764.3445066},
    urldate = {2025-08-03},
    booktitle = {Proceedings of the 2021 {CHI} {Conference} on {Human} {Factors} in {Computing} {Systems}},
    publisher = {Association for Computing Machinery},
    author = {Krosnick, Rebecca and Anderson, Fraser and Matejka, Justin and Oney, Steve and S. Lasecki, Walter and Grossman, Tovi and Fitzmaurice, George},
    month = may,
    year = {2021},
    pages = {1--13},
}

@inproceedings{yun_generative_2025,
    address = {New York, NY, USA},
    series = {{CHI} '25},
    title = {Generative {AI} in {Knowledge} {Work}: {Design} {Implications} for {Data} {Navigation} and {Decision}-{Making}},
    isbn = {979-8-4007-1394-1},
    shorttitle = {Generative {AI} in {Knowledge} {Work}},
    url = {https://doi.org/10.1145/3706598.3713337},
    doi = {10.1145/3706598.3713337},
    urldate = {2025-08-11},
    booktitle = {Proceedings of the 2025 {CHI} {Conference} on {Human} {Factors} in {Computing} {Systems}},
    publisher = {Association for Computing Machinery},
    author = {Yun, Bhada and Feng, Dana and Chen, Ace S. and Nikzad, Afshin and Salehi, Niloufar},
    month = apr,
    year = {2025},
    pages = {1--19},
}

@misc{noauthor_v0_nodate,
    title = {v0 by {Vercel}},
    url = {https://v0.app/},
    language = {en},
    urldate = {2025-08-14},
}

@misc{noauthor_claude_nodate,
    title = {Claude 3.7 {Sonnet} and {Claude} {Code}},
    url = {https://www.anthropic.com/news/claude-3-7-sonnet},
    language = {en},
    urldate = {2025-04-14},
}

@article{braun_using_2006,
    title = {Using thematic analysis in psychology},
    volume = {3},
    issn = {1478-0887},
    url = {https://doi.org/10.1191/1478088706qp063oa},
    doi = {10.1191/1478088706qp063oa},
    number = {2},
    urldate = {2025-08-24},
    journal = {Qualitative Research in Psychology},
    author = {Braun, Virginia and Clarke, Victoria},
    month = jan,
    year = {2006},
    note = {Publisher: Routledge
\_eprint: https://doi.org/10.1191/1478088706qp063oa},
    pages = {77--101},
}

@misc{noauthor_cursor_nodate,
    title = {Cursor - {The} {AI} {Code} {Editor}},
    url = {https://cursor.com/home},
    urldate = {2025-08-24},
}

@inproceedings{kasica_dirty_2023,
    address = {New York, NY, USA},
    series = {{CHI} '23},
    title = {Dirty {Data} in the {Newsroom}: {Comparing} {Data} {Preparation} in {Journalism} and {Data} {Science}},
    isbn = {978-1-4503-9421-5},
    shorttitle = {Dirty {Data} in the {Newsroom}},
    url = {https://doi.org/10.1145/3544548.3581271},
    doi = {10.1145/3544548.3581271},
    urldate = {2025-08-24},
    booktitle = {Proceedings of the 2023 {CHI} {Conference} on {Human} {Factors} in {Computing} {Systems}},
    publisher = {Association for Computing Machinery},
    author = {Kasica, Stephen and Berret, Charles and Munzner, Tamara},
    month = apr,
    year = {2023},
    pages = {1--18},
}

@misc{chen_generative_2025,
    title = {Generative {Interfaces} for {Language} {Models}},
    url = {http://arxiv.org/abs/2508.19227},
    doi = {10.48550/arXiv.2508.19227},
    urldate = {2025-08-29},
    publisher = {arXiv},
    author = {Chen, Jiaqi and Zhang, Yanzhe and Zhang, Yutong and Shao, Yijia and Yang, Diyi},
    month = aug,
    year = {2025},
    note = {arXiv:2508.19227 [cs]},
}

@inproceedings{kandel_wrangler_2011,
    address = {New York, NY, USA},
    series = {{CHI} '11},
    title = {Wrangler: interactive visual specification of data transformation scripts},
    isbn = {978-1-4503-0228-9},
    shorttitle = {Wrangler},
    url = {https://dl.acm.org/doi/10.1145/1978942.1979444},
    doi = {10.1145/1978942.1979444},
    urldate = {2025-09-08},
    booktitle = {Proceedings of the {SIGCHI} {Conference} on {Human} {Factors} in {Computing} {Systems}},
    publisher = {Association for Computing Machinery},
    author = {Kandel, Sean and Paepcke, Andreas and Hellerstein, Joseph and Heer, Jeffrey},
    month = may,
    year = {2011},
    pages = {3363--3372},
}

@misc{noauthor_data_nodate,
    title = {Data analysis with {ChatGPT}},
    url = {https://help.openai.com/en/articles/8437071-data-analysis-with-chatgpt},
    language = {en},
    urldate = {2025-09-10},
    journal = {OpenAI Help Center},
}

@inproceedings{chen_empathy-based_2024,
    address = {Honolulu HI USA},
    title = {An {Empathy}-{Based} {Sandbox} {Approach} to {Bridge} the {Privacy} {Gap} among {Attitudes}, {Goals}, {Knowledge}, and {Behaviors}},
    isbn = {979-8-4007-0330-0},
    url = {https://dl.acm.org/doi/10.1145/3613904.3642363},
    doi = {10.1145/3613904.3642363},
    language = {en},
    urldate = {2025-09-11},
    booktitle = {Proceedings of the {CHI} {Conference} on {Human} {Factors} in {Computing} {Systems}},
    publisher = {ACM},
    author = {Chen, Chaoran and Li, Weijun and Song, Wenxin and Ye, Yanfang and Yao, Yaxing and Li, Toby Jia-Jun},
    month = may,
    year = {2024},
    pages = {1--28},
}

@misc{chen_why_2025,
    title = {Why am {I} seeing this: {Democratizing} {End} {User} {Auditing} for {Online} {Content} {Recommendations}},
    shorttitle = {Why am {I} seeing this},
    url = {http://arxiv.org/abs/2410.04917},
    doi = {10.48550/arXiv.2410.04917},
    urldate = {2025-09-12},
    publisher = {arXiv},
    author = {Chen, Chaoran and Li, Leyang and Cao, Luke and Ye, Yanfang and Li, Tianshi and Yao, Yaxing and Li, Toby Jia-jun},
    month = apr,
    year = {2025},
    note = {arXiv:2410.04917 [cs]},
}

@misc{noauthor_replit_nodate,
    title = {Replit – {Build} apps and sites with {AI}},
    url = {https://replit.com/},
    urldate = {2025-09-12},
}

@misc{noauthor_potluck_nodate,
    title = {Potluck: {Dynamic} documents as personal software},
    shorttitle = {Potluck},
    url = {https://www.inkandswitch.com/potluck/},
    language = {en},
    urldate = {2026-01-08},
}

@inproceedings{drosos_dynamic_2025,
    address = {Amsterdam Netherlands},
    title = {Dynamic {Prompt} {Middleware}: {Contextual} {Prompt} {Refinement} {Controls} for {Comprehension} {Tasks}},
    isbn = {979-8-4007-1384-2},
    shorttitle = {Dynamic {Prompt} {Middleware}},
    url = {https://dl.acm.org/doi/10.1145/3729176.3729203},
    doi = {10.1145/3729176.3729203},
    language = {en},
    urldate = {2025-12-16},
    booktitle = {Proceedings of the 4th {Annual} {Symposium} on {Human}-{Computer} {Interaction} for {Work}},
    publisher = {ACM},
    author = {Drosos, Ian and Williams, Jack and Sarkar, Advait and Wilson, Nicholas and Rintel, Sean and Panda, Payod},
    month = jun,
    year = {2025},
    pages = {1--23},
}

@misc{noauthor_figma_nodate,
    title = {Figma {Make}: {Create} with {AI}-{Powered} {Design} {Tools}},
    shorttitle = {Figma {Make}},
    url = {https://www.figma.com/make/},
    language = {en},
    urldate = {2026-01-08},
    journal = {Figma},
}

@misc{noauthor_introducing_nodate,
    title = {Introducing {A2UI}: {An} open project for agent-driven interfaces- {Google} {Developers} {Blog}},
    shorttitle = {Introducing {A2UI}},
    url = {https://developers.googleblog.com/introducing-a2ui-an-open-project-for-agent-driven-interfaces/},
    language = {en},
    urldate = {2026-01-08},
}

@inproceedings{min_meridian_2025,
    address = {New York, NY, USA},
    series = {{UIST} '25},
    title = {Meridian: {A} {Design} {Framework} for {Malleable} {Overview}-{Detail} {Interfaces}},
    isbn = {979-8-4007-2037-6},
    shorttitle = {Meridian},
    url = {https://dl.acm.org/doi/10.1145/3746059.3747654},
    doi = {10.1145/3746059.3747654},
    urldate = {2026-01-08},
    booktitle = {Proceedings of the 38th {Annual} {ACM} {Symposium} on {User} {Interface} {Software} and {Technology}},
    publisher = {Association for Computing Machinery},
    author = {Min, Bryan and Xia, Haijun},
    month = sep,
    year = {2025},
    pages = {1--14},
}

@inproceedings{shankar_steering_2025,
    address = {Busan Republic of Korea},
    title = {Steering {Semantic} {Data} {Processing} {With} {DocWrangler}},
    isbn = {979-8-4007-2037-6},
    url = {https://dl.acm.org/doi/10.1145/3746059.3747625},
    doi = {10.1145/3746059.3747625},
    language = {en},
    urldate = {2026-01-09},
    booktitle = {Proceedings of the 38th {Annual} {ACM} {Symposium} on {User} {Interface} {Software} and {Technology}},
    publisher = {ACM},
    author = {Shankar, Shreya and Chopra, Bhavya and Hasan, Mawil and Lee, Stephen and Hartmann, Bjoern and Hellerstein, Joseph and Parameswaran, Aditya and Wu, Eugene},
    month = sep,
    year = {2025},
    pages = {1--18},
}

@inproceedings{freund_flowco_2025,
    address = {Busan Republic of Korea},
    title = {Flowco: {Mixed}-{Initiative} {Authoring} of {Reliable} {End}-to-{End} {Data} {Analyses} via {Dataflow} {Graphs} and {LLMs}},
    isbn = {979-8-4007-2037-6},
    shorttitle = {Flowco},
    url = {https://dl.acm.org/doi/10.1145/3746059.3747636},
    doi = {10.1145/3746059.3747636},
    language = {en},
    urldate = {2026-01-06},
    booktitle = {Proceedings of the 38th {Annual} {ACM} {Symposium} on {User} {Interface} {Software} and {Technology}},
    publisher = {ACM},
    author = {Freund, Stephen N. and Simon, Brooke and Berger, Emery D. and Jun, Eunice},
    month = sep,
    year = {2025},
    pages = {1--20},
}

@inproceedings{gu_how_2024,
    address = {New York, NY, USA},
    series = {{CHI} '24},
    title = {How {Do} {Analysts} {Understand} and {Verify} {AI}-{Assisted} {Data} {Analyses}?},
    isbn = {979-8-4007-0330-0},
    url = {https://dl.acm.org/doi/10.1145/3613904.3642497},
    doi = {10.1145/3613904.3642497},
    urldate = {2026-01-10},
    booktitle = {Proceedings of the 2024 {CHI} {Conference} on {Human} {Factors} in {Computing} {Systems}},
    publisher = {Association for Computing Machinery},
    author = {Gu, Ken and Shang, Ruoxi and Althoff, Tim and Wang, Chenglong and Drucker, Steven M.},
    month = may,
    year = {2024},
    pages = {1--22},
}

@misc{payandeh_noteex_2025,
    title = {{NoteEx}: {Interactive} {Visual} {Context} {Manipulation} for {LLM}-{Assisted} {Exploratory} {Data} {Analysis} in {Computational} {Notebooks}},
    shorttitle = {{NoteEx}},
    url = {http://arxiv.org/abs/2511.07223},
    doi = {10.48550/arXiv.2511.07223},
    urldate = {2026-01-12},
    publisher = {arXiv},
    author = {Payandeh, Mohammad Hasan and Yuan, Lin-Ping and Zhao, Jian},
    month = nov,
    year = {2025},
    note = {arXiv:2511.07223 [cs]},
}

@article{weng_insightlens_2025,
    title = {{InsightLens}: {Augmenting} {LLM}-{Powered} {Data} {Analysis} {With} {Interactive} {Insight} {Management} and {Navigation}},
    volume = {31},
    issn = {1941-0506},
    shorttitle = {{InsightLens}},
    url = {https://ieeexplore.ieee.org/abstract/document/10989518},
    doi = {10.1109/TVCG.2025.3567131},
    number = {6},
    urldate = {2026-01-11},
    journal = {IEEE Transactions on Visualization and Computer Graphics},
    author = {Weng, Luoxuan and Wang, Xingbo and Lu, Junyu and Feng, Yingchaojie and Liu, Yihan and Feng, Haozhe and Huang, Danqing and Chen, Wei},
    month = jun,
    year = {2025},
    pages = {3719--3732},
}

@misc{dellacqua_navigating_2023,
    address = {Rochester, NY},
    type = {{SSRN} {Scholarly} {Paper}},
    title = {Navigating the {Jagged} {Technological} {Frontier}: {Field} {Experimental} {Evidence} of the {Effects} of {AI} on {Knowledge} {Worker} {Productivity} and {Quality}},
    shorttitle = {Navigating the {Jagged} {Technological} {Frontier}},
    url = {https://papers.ssrn.com/abstract=4573321},
    doi = {10.2139/ssrn.4573321},
    language = {en},
    urldate = {2026-01-12},
    publisher = {Social Science Research Network},
    author = {Dell'Acqua, Fabrizio and McFowland III, Edward and Mollick, Ethan R. and Lifshitz-Assaf, Hila and Kellogg, Katherine and Rajendran, Saran and Krayer, Lisa and Candelon, François and Lakhani, Karim R.},
    month = sep,
    year = {2023},
}

@inproceedings{ma_proactiveagent_2023,
    address = {New York, NY, USA},
    series = {{UIST} '23 {Adjunct}},
    title = {{ProactiveAgent}: {Personalized} {Context}-{Aware} {Reminder} {System}},
    isbn = {979-8-4007-0096-5},
    shorttitle = {{ProactiveAgent}},
    url = {https://dl.acm.org/doi/10.1145/3586182.3625115},
    doi = {10.1145/3586182.3625115},
    urldate = {2026-01-11},
    booktitle = {Adjunct {Proceedings} of the 36th {Annual} {ACM} {Symposium} on {User} {Interface} {Software} and {Technology}},
    publisher = {Association for Computing Machinery},
    author = {Ma, Yumeng and Ren, Jiahao},
    month = oct,
    year = {2023},
    pages = {1--3},
}

@misc{noauthor_user_nodate,
    title = {User {Experience} {Design} {Professionals}’ {Perceptions} of {Generative} {Artificial} {Intelligence} {\textbar} {Proceedings} of the 2024 {CHI} {Conference} on {Human} {Factors} in {Computing} {Systems}},
    url = {https://dl.acm.org/doi/full/10.1145/3613904.3642114},
    urldate = {2026-01-12},
}

@inproceedings{zhang_rethinking_2024,
    address = {New York, NY, USA},
    series = {{CHI} '24},
    title = {Rethinking {Human}-{AI} {Collaboration} in {Complex} {Medical} {Decision} {Making}: {A} {Case} {Study} in {Sepsis} {Diagnosis}},
    isbn = {979-8-4007-0330-0},
    shorttitle = {Rethinking {Human}-{AI} {Collaboration} in {Complex} {Medical} {Decision} {Making}},
    url = {https://dl.acm.org/doi/10.1145/3613904.3642343},
    doi = {10.1145/3613904.3642343},
    urldate = {2026-01-11},
    booktitle = {Proceedings of the 2024 {CHI} {Conference} on {Human} {Factors} in {Computing} {Systems}},
    publisher = {Association for Computing Machinery},
    author = {Zhang, Shao and Yu, Jianing and Xu, Xuhai and Yin, Changchang and Lu, Yuxuan and Yao, Bingsheng and Tory, Melanie and Padilla, Lace M. and Caterino, Jeffrey and Zhang, Ping and Wang, Dakuo},
    month = may,
    year = {2024},
    pages = {1--18},
}

@inproceedings{gao_collabcoder_2024,
    address = {Honolulu HI USA},
    title = {{CollabCoder}: {A} {Lower}-barrier, {Rigorous} {Workflow} for {Inductive} {Collaborative} {Qualitative} {Analysis} with {Large} {Language} {Models}},
    isbn = {979-8-4007-0330-0},
    shorttitle = {{CollabCoder}},
    url = {https://dl.acm.org/doi/10.1145/3613904.3642002},
    doi = {10.1145/3613904.3642002},
    language = {en},
    urldate = {2026-01-07},
    booktitle = {Proceedings of the {CHI} {Conference} on {Human} {Factors} in {Computing} {Systems}},
    publisher = {ACM},
    author = {Gao, Jie and Guo, Yuchen and Lim, Gionnieve and Zhang, Tianqin and Zhang, Zheng and Li, Toby Jia-Jun and Perrault, Simon Tangi},
    month = may,
    year = {2024},
    pages = {1--29},
}

@inproceedings{aveni_generative_2025,
    address = {Busan Republic of Korea},
    title = {Generative {Trigger}-{Action} {Programming} with {Ply}},
    isbn = {979-8-4007-2037-6},
    url = {https://dl.acm.org/doi/10.1145/3746059.3747638},
    doi = {10.1145/3746059.3747638},
    language = {en},
    urldate = {2025-11-10},
    booktitle = {Proceedings of the 38th {Annual} {ACM} {Symposium} on {User} {Interface} {Software} and {Technology}},
    publisher = {ACM},
    author = {Aveni, Timothy J. and Mor, Hila and Fox, Armando and Hartmann, Björn},
    month = sep,
    year = {2025},
    pages = {1--17},
}

@inproceedings{rule_exploration_2018,
    address = {Montreal QC Canada},
    title = {Exploration and {Explanation} in {Computational} {Notebooks}},
    isbn = {978-1-4503-5620-6},
    url = {https://dl.acm.org/doi/10.1145/3173574.3173606},
    doi = {10.1145/3173574.3173606},
    language = {en},
    urldate = {2026-01-15},
    booktitle = {Proceedings of the 2018 {CHI} {Conference} on {Human} {Factors} in {Computing} {Systems}},
    publisher = {ACM},
    author = {Rule, Adam and Tabard, Aurélien and Hollan, James D.},
    month = apr,
    year = {2018},
    pages = {1--12},
}

@article{gadhave_persist_2024,
    title = {Persist: {Persistent} and {Reusable} {Interactions} in {Computational} {Notebooks}},
    language = {en},
    author = {Gadhave, K and Cutler, Z and Lex, A},
    year = {2024},
}

@inproceedings{mackay_patterns_1990,
    address = {New York, NY, USA},
    series = {{CSCW} '90},
    title = {Patterns of sharing customizable software},
    isbn = {978-0-89791-402-4},
    url = {https://dl.acm.org/doi/10.1145/99332.99356},
    doi = {10.1145/99332.99356},
    urldate = {2026-01-16},
    booktitle = {Proceedings of the 1990 {ACM} conference on {Computer}-supported cooperative work},
    publisher = {Association for Computing Machinery},
    author = {Mackay, Wendy E.},
    month = sep,
    year = {1990},
    pages = {209--221},
}

@article{deacon_model-view-controller_nodate,
    title = {Model-{View}-{Controller} ({MVC}) {Architecture}},
    language = {en},
    author = {Deacon, John},
}

@inproceedings{fok_marco_2024,
    address = {New York, NY, USA},
    series = {{CHI} '24},
    title = {Marco: {Supporting} {Business} {Document} {Workflows} via {Collection}-{Centric} {Information} {Foraging} with {Large} {Language} {Models}},
    isbn = {979-8-4007-0330-0},
    shorttitle = {Marco},
    url = {https://dl.acm.org/doi/10.1145/3613904.3641969},
    doi = {10.1145/3613904.3641969},
    urldate = {2026-01-20},
    booktitle = {Proceedings of the 2024 {CHI} {Conference} on {Human} {Factors} in {Computing} {Systems}},
    publisher = {Association for Computing Machinery},
    author = {Fok, Raymond and Lipka, Nedim and Sun, Tong and Siu, Alexa F},
    month = may,
    year = {2024},
    pages = {1--20},
}

@misc{noauthor_passages_nodate,
    title = {Passages: {Interacting} with {Text} {Across} {Documents} {\textbar} {Proceedings} of the 2022 {CHI} {Conference} on {Human} {Factors} in {Computing} {Systems}},
    url = {https://dl.acm.org/doi/abs/10.1145/3491102.3502052?utm_source=chatgpt.com},
    urldate = {2026-01-20},
}

@misc{noauthor_managing_nodate,
    title = {Managing {Messes} in {Computational} {Notebooks} {\textbar} {Proceedings} of the 2019 {CHI} {Conference} on {Human} {Factors} in {Computing} {Systems}},
    url = {https://dl.acm.org/doi/10.1145/3290605.3300500?utm_source=chatgpt.com},
    urldate = {2026-01-20},
}

@inproceedings{ma_ambigchat_2025,
    address = {Busan Republic of Korea},
    title = {{AmbigChat}: {Interactive} {Hierarchical} {Clarification} for {Ambiguous} {Open}-{Domain} {Question} {Answering}},
    isbn = {979-8-4007-2037-6},
    shorttitle = {{AmbigChat}},
    url = {https://dl.acm.org/doi/10.1145/3746059.3747686},
    doi = {10.1145/3746059.3747686},
    language = {en},
    urldate = {2025-12-16},
    booktitle = {Proceedings of the 38th {Annual} {ACM} {Symposium} on {User} {Interface} {Software} and {Technology}},
    publisher = {ACM},
    author = {Ma, Jiaju and Shi, Lei and Robertsen, Kenneth Aleksander and Chi, Peggy},
    month = sep,
    year = {2025},
    pages = {1--18},
}

@inproceedings{leung_squire_2025,
    address = {Busan Republic of Korea},
    title = {{SQUIRE}: {Interactive} {UI} {Authoring} via {Slot} {QUery} {Intermediate} {REpresentations}},
    isbn = {979-8-4007-2037-6},
    shorttitle = {{SQUIRE}},
    url = {https://dl.acm.org/doi/10.1145/3746059.3747672},
    doi = {10.1145/3746059.3747672},
    language = {en},
    urldate = {2026-01-20},
    booktitle = {Proceedings of the 38th {Annual} {ACM} {Symposium} on {User} {Interface} {Software} and {Technology}},
    publisher = {ACM},
    author = {Leung, Alan and Cheng, Ruijia and Wu, Jason and Nichols, Jeffrey and Barik, Titus},
    month = sep,
    year = {2025},
    pages = {1--17},
}

@inproceedings{suh_storyensemble_2025,
    address = {Busan Republic of Korea},
    title = {{StoryEnsemble}: {Enabling} {Dynamic} {Exploration} \& {Iteration} in the {Design} {Process} with {AI} and {Forward}-{Backward} {Propagation}},
    isbn = {979-8-4007-2037-6},
    shorttitle = {{StoryEnsemble}},
    url = {https://dl.acm.org/doi/10.1145/3746059.3747772},
    doi = {10.1145/3746059.3747772},
    language = {en},
    urldate = {2026-01-20},
    booktitle = {Proceedings of the 38th {Annual} {ACM} {Symposium} on {User} {Interface} {Software} and {Technology}},
    publisher = {ACM},
    author = {Suh, Sangho and Lai, Michael and Pu, Kevin and Dow, Steven P. and Grossman, Tovi},
    month = sep,
    year = {2025},
    pages = {1--36},
}

@misc{noauthor_graffiti_nodate,
    title = {Graffiti: {Enabling} an {Ecosystem} of {Personalized} and {Interoperable} {Social} {Applications} {\textbar} {Proceedings} of the 38th {Annual} {ACM} {Symposium} on {User} {Interface} {Software} and {Technology}},
    url = {https://dl.acm.org/doi/10.1145/3746059.3747627},
    urldate = {2026-01-20},
}

@misc{tang_naturaledit_2026,
    title = {{NaturalEdit}: {Code} {Modification} through {Direct} {Interaction} with {Adaptive} {Natural} {Language} {Representation}},
    shorttitle = {{NaturalEdit}},
    url = {http://arxiv.org/abs/2510.04494},
    doi = {10.48550/arXiv.2510.04494},
    urldate = {2026-07-21},
    publisher = {arXiv},
    author = {Tang, Ningzhi and Meininger, David and Xu, Gelei and Shi, Yiyu and Huang, Yu and McMillan, Collin and Li, Toby Jia-Jun},
    month = apr,
    year = {2026},
    note = {arXiv:2510.04494 [cs.HC]},
}


\end{document}